\documentclass[journal=jacsat,manuscript=article]{achemso}
\usepackage[version=3]{mhchem}

\usepackage{adjustbox}
\usepackage{graphicx}
\usepackage{dcolumn}
\usepackage{bm}
\usepackage{amsmath}            
\usepackage{amsfonts}           
\usepackage{amssymb}
\usepackage{array}
\usepackage{multirow}
\usepackage{graphicx}
\usepackage{subfigure} 
\usepackage{epstopdf}
\usepackage{lineno}
\usepackage{booktabs}
\usepackage[flushleft]{threeparttable}
\usepackage{xcolor}
\usepackage{color,soul}
\usepackage[english]{babel}
\usepackage{polski}
\AppendGraphicsExtensions{.tif}
\usepackage{placeins} 
\title{Half-metallic to ferromagnetic phase transition in CrSH monolayer using DFT+$U$ and BO-MD calculations}

\author{Akkarach Sukserm}
\affiliation{Department of Physics, Faculty of Science, Chulalongkorn University, Bangkok,10330, Thailand}
\author{Jakkapat Seeyangnok}
\author{Udomsilp Pinsook}
\email{Udomsilp.P@Chula.ac.th}
\affiliation{Department of Physics, Faculty of Science, Chulalongkorn University, Bangkok,10330, Thailand}

\begin{document}
 \begin{abstract}
We present a comprehensive investigation of the structural, electronic, magnetic, and vibrational properties of CrSH monolayers in the 1T and 2H phases using density functional theory (DFT)+$U$ calculations with a converged Hubbard $U$ value of 5.54 eV and Born-Oppenheimer molecular dynamics (BO-MD) simulations. The ferromagnetic (FM) 1T-CrSH phase is found to be dynamically and thermodynamically stable, exhibiting semiconducting behavior with a band gap of 1.1 eV and a magnetic moment of 3.0 $\mu_B$ per Cr atom. On the other hand, the 2H-CrSH phase is a half-metallic (HM) phase. We found that it is a metastable phase and undergoes a rapid phase transition to the 1T phase under finite temperature at 300 K. Phonon calculations, performed using the finite displacement method and corrected for rotational invariance corrections with Huang and Born-Huang sum rules, resolve spurious imaginary frequencies in the flexural ZA phonon mode near the $\Gamma$-point, ensuring physical accuracy. These findings establish CrSH monolayers as promising candidates for spintronic and valleytronic applications, with tunable electronic properties enabled by phase engineering.
\end{abstract}



\section{Introduction} \label{Sec:Intro}

In recent years, two-dimensional (2D) materials have revolutionized materials science, offering exceptional physical properties and enabling a wide range of technological applications, including nanoelectronics\cite{nanoelect_2015}, optoelectronics\cite{MoS2_2001,MoS2_Optoelec2019,Optoelectronics2022}, energy storage\cite{storage2020,storage2021}, and catalysis\cite{WS2MoS2_2004,MoS2_App2019}. Among the vast family of 2D materials, transition metal dichalcogenides (TMDs), such as MoS$_2$, MoSe$_2$, WS$_2$, and WSe$_2$, stand out for their tunable electronic, magnetic, and vibrational properties. These characteristics make them promising candidates for next-generation flexible and transparent devices \cite{MoS2_2001,WS2MoS2_2004, MoS2-1_2012,MoS2_2014,MoS2_HM_2014,MoS2_App2019,MoS2_Optoelec2019,MoSe2_2020,MoSe2_2022}.

The discovery of distinct polymorphic phases in TMDs, such as 1T and 2H phases, has further expanded the scope of their applications by introducing phase engineering—a strategy to tailor material properties by manipulating their atomic arrangements. For example, the coexistence of these phases in MoS$_2$ has sparked intense research into phase transitions, which offer new opportunities to enhance device performance. Beyond their structural diversity, magnetic 2D materials have emerged as a compelling research frontier, exhibiting a variety of phases, including ferromagnetic (FM), antiferromagnetic (AFM), and half-metallic (HM) states, which hold significant promise for spintronic and valleytronic applications \cite{MoS2-2_2012,valley2016,MoS2_HM_2015,MoS2_2016,MoS2_HM_2017,MoS2_HM_2018,MoS2_AFM_2021}. 

Several examples highlight the versatility of TMDs. Monolayer VS$_2$, with its intrinsic FM behavior, has been identified as a leading candidate for spintronic devices \cite{VS2_2013,VS2_2017}. The CoCl$_2$ and CoBr$_2$ monolayers are FM semiconductors \cite{CoCl2_2018,CoBr2_2019,CoCl2_2023}, while MnS$_2$ and MnSe$_2$ exhibit narrow-gap FM semiconducting properties \cite{MnS2_2014}. Recently, Janus structures such as MoSSe \cite{Spin_2020,lu2017janus}, WSSe\cite{Spin_2020}, and PtSSe \cite{sant2020synthesis}, which introduce atomic asymmetry by replacing atoms on one side of the layer, have demonstrated unique physical properties that enable precise control over electronic and magnetic characteristics. For instance, Janus structures like Cr$_2$I$_3$X$_3$ (X = Br, Cl) and VSSe have been reported as FM semiconductors, while FeXY (X, Y = Cl, Br, I) and V$_2$XN (X = P, As) exhibit FM metallic behavior \cite{VSSe_2019}. 
One of the most exciting developments in this field is the ability to enhance the Curie temperature ($T_c$) of these materials through structural engineering. While VI$_3$ and FeI$_3$ have a low $T_c$ of 50 K and 77 K, respectively \cite{VI3-1_2019,VI3-2_2019,Fe2Cl3I3_2020}, Janus structures like Cr$_2$Cl$_3$I$_3$, V$_2$Cl$_3$I$_3$ and Fe$_2$I$_3$Cl$_3$ exhibit $T_c$ values of 180 K, 240 K, and 260 K, respectively \cite{Cr2Cl3I3_2024,V2Cl3I3_2020,Fe2Cl3I3_2020,FeClF_2023}. These advancements underscore the growing potential of Janus 2D materials to achieve room-temperature magnetism, a critical milestone for their integration into spintronic and valleytronic devices \cite{Spin_2017,Spin_2020,Valley_2023}. 

Building upon recent breakthroughs, the synthesis of Janus MoSH monolayers through H$_2$-plasma treatment has opened new pathways for designing hydrogenated transition metal chalcogenides \cite{Spin_2020,tang20222d}. By selectively replacing top-layer sulfur atoms in MoS$_2$ with hydrogen, researchers have demonstrated the feasibility of creating dynamically stable structures with unique properties. For instance, MoSH exhibits two-gap superconductivity with a transition temperature of 28.58 K \cite{liu2022two,ku2023ab}, sparking interest in extending the concept of hydrogenation to other transition metal chalcogenides \cite{seeyangnok2024superconductivity,li2024machine,ul2024superconductivity,jseeyangPRBWXH2024,jseeyangnok2024JTMCS}. 

Among these innovations, the Janus monolayer CrSH in the 1T phase (1T-CrSH) has emerged as a promising candidate for spintronic applications \cite{lu2017janus}. Initial studies reveal that 1T-CrSH exhibits a ferromagnetic phase with a magnetic moment of 3.0 $\mu_B$ per Cr atom, predominantly arising from Cr(3$d$)-orbitals. Its Néel temperature ($T_N$), estimated at 193 K through Monte Carlo simulations, can be further enhanced to 402 K under a 5$\%$ tensile strain \cite{CrSH_1T_2024}, demonstrating its tunability for advanced applications. Additionally, the heterostructure of 1T-CrSH with WS$_2$ shows significant valley splitting in both conduction and valence bands, indicating it as a strong contender for valleytronic technologies.

Despite these promising properties, the accurate description of stability, phase transitions, and physical properties of CrSH monolayer poses significant computational and methodological challenges. 
Density functional theory (DFT) often requires the on-site Hubbard $U$ corrections (DFT+$U$) to describe the strongly correlated 3$d$-orbitals of transition metal atoms. Furthermore, vibrational property calculations for 2D materials demand careful consideration of lattice invariance conditions, as improper handling can lead to unphysical artifacts, such as spurious imaginary frequencies in the flexural ZA phonon mode near the $\Gamma$-point \cite{ZA_2015,hiphive2019,hiphive2020}. These challenges underscore the need for advanced computational approaches to ensure reliable predictions of CrSH properties.

This study investigates the structural, electronic, magnetic, and vibrational properties of CrSH monolayers in the 1T and 2H phases using the \textit{ab}-initio DFT+$U$ calculations. The structural stability and phase transition pathways were explored using total energy calculations and Born-Oppenheimer molecular dynamics (BO-MD) simulations at 300 K within the NVT ensemble, maintaining a constant number of particles (N), volume (V), and temperature (T). This methodology ensures a robust dynamic and thermodynamic stability assessment under realistic conditions. 

The electronic and magnetic properties were explored and analyzed to understand the semiconducting and spintronic potential, with particular attention to the role of Cr($3d$)-orbital interactions in driving magnetic behavior. The vibrational properties, including the phonon dispersion and the phonon density of states (Phonon DOS) were computed using the finite displacement method combined with the rotational invariance corrections through Huang and Born-Huang sum rules to ensure physically accurate results \cite{bornhuang1996,hiphive2019,hiphive2020}. 

The findings reveal the FM 1T-CrSH phase is a dynamically and thermodynamically stable structure, exhibiting semiconducting behavior with a band gap of 1.1 eV and a magnetic moment of 3.0 $\mu_B$ per Cr atom. In contrast, the FM 2H-CrSH phase is metastable, undergoing a rapid phase transition at 300K and exhibiting an HM behavior during the transformation. These insights would position the CrSH monolayer as a promising candidate for next-generation 2D material-based devices, with potential applications in spintronics, valleytronics, and phase-engineered electronics. 

\begin{figure*}[tbh!]
  \centering
 \includegraphics[width=0.85\linewidth]{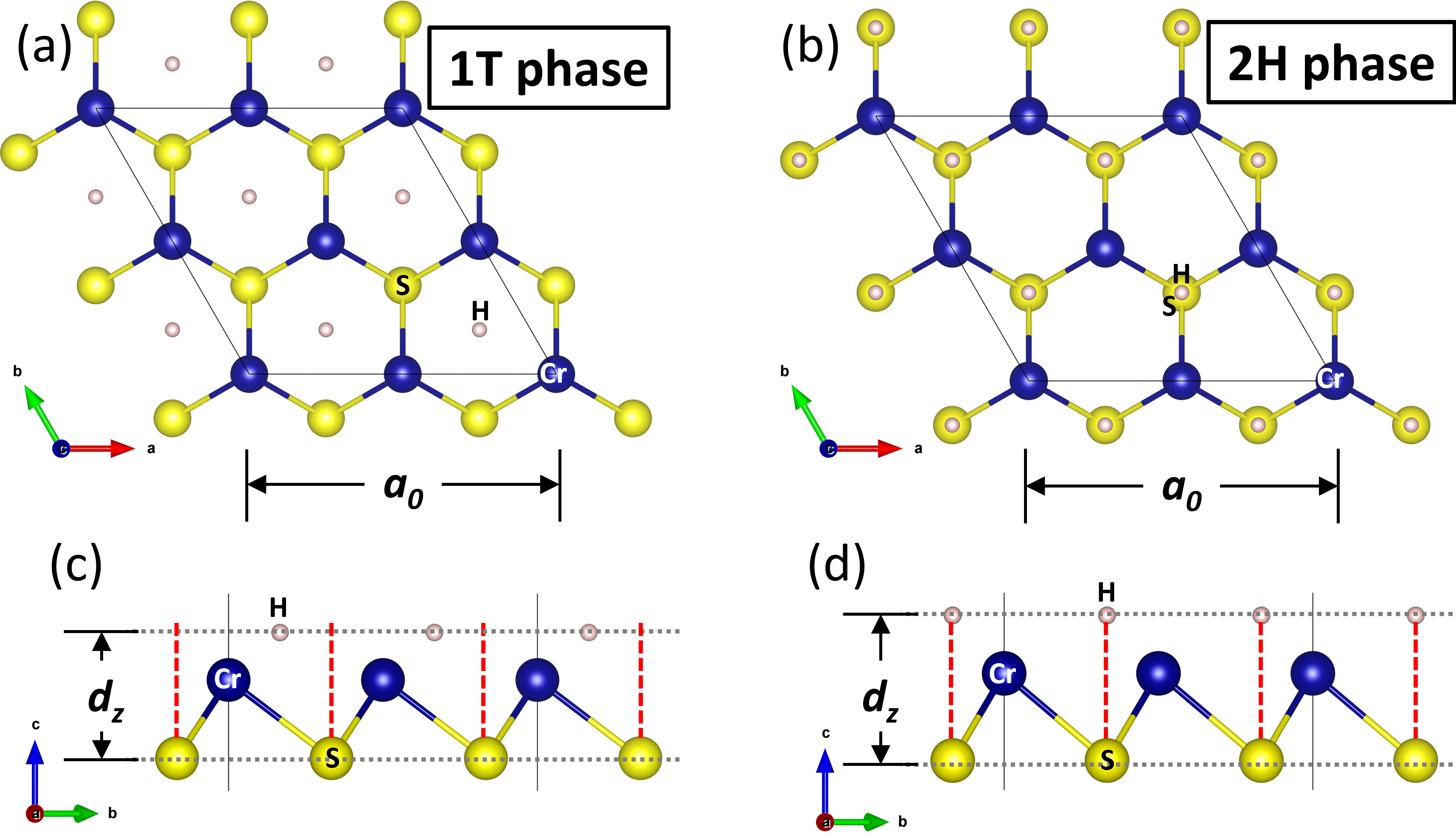}
\caption{Crystal structures of CrSH monolayer, (a) 1T-phase Top view, (b) 2H-phase Top view, (c) 1T-phase Side view, and (d) 2H-phase Side view.} 
\label{fig:structure}%
\end{figure*}

\section{Computational details} \label{Sec:Theory}

We performed a comprehensive set of the DFT calculations to investigate the structural, electronic, magnetic, and vibrational properties of both 1T-CrSH and 2H-CrSH monolayers, as shown in Figures~\ref{fig:structure}(a) and~\ref{fig:structure}(b), respectively. These calculations were carried out using the Quantum-ESPRESSO suite \cite{DFT1964,QE-2017,QE-2020}, which enabled detailed modeling of the crystal structures and electronic behaviors of these 2D materials. The study focused on 12-atom unit cells (2 x 2 x 1 supercells of CrSH) for both phases. 
To account for the strong on-site Hubbard correlations in the Cr($3d$)-orbitals, the DFT+$U$ approach was applied \cite{DFT_U2005,timrov2022hp}. The Hubbard $U$ value was determined via linear response theory (LR-DFT) as $U = (\chi^{-1}_0 - \chi^{-1})_{II}$, where $\chi_0$ and $\chi$ are the bare and interacting response matrices, respectively. The DFPT approach \cite{timrov2018hubbard,timrov2021self} has improved to compute the response matrix $\chi_{IJ}$ by summing contributions from various
$q$-mesh grids over the Brillouin zone, 
\begin{equation}
    \begin{aligned}
    \chi_{IJ} = \frac{1}{N_q} \sum_q^{N_q} e^{i\textbf{q} \cdot (\textbf{R}_I -\textbf{R}_J)} \cdot \frac{1}{N_k} \sum_k^{N_k} \sum_i^{N_e} \left[ \langle u_{i,\textbf{k}} | \varphi^I_{m_1,\textbf{k}} \rangle \langle \varphi^I_{m_2,\textbf{k+q}} |\Delta^J_{\textbf{q}} u_{i,\textbf{k}}  \rangle \right] ,
        \end{aligned}
    \label{eq:DFPT}
\end{equation}
where $N_e$ is the number of occupied electrons, $N_k$ and $N_q$ are the number of $k$-and $q$-points in the Brillouin zone, and $\Delta^J_q$ represents the perturbation term. In this study, $q$-meshes ranging from 2×2×1 to 14×14×1 were evaluated to converge the response matrices, corresponding to supercell sizes required in the LR-DFT approach \cite{DFT_U2005,timrov2018hubbard}. The Hubbard $U$ value for Cr was converged at 5.52 eV. The exchange-correlation energy was treated using the Perdew-Burke-Ernzerhof (PBE) \cite{perdew1996generalized} generalized gradient approximation (GGA), including spin polarization to capture magnetic properties. 

For the self-consistent field (SCF) calculations, structural optimization, and electronic properties determination, we used the projector augmented-wave (PAW) pseudopotentials \cite{kresse1999ultrasoft,hobbs2000fully} for the Cr ($3s^2~4s^2~3p^6~3d^4$), S ($3s^2 ~3p^4$), and H ($1s^1$) atoms. A plane-wave energy cut-off of 80 Ry was chosen to ensure accurate energy convergence, and the Brillouin zone was sampled using a dense 14×14×2 Monkhorst-Pack k-point grid \cite{monkhorst1976special} to achieve energy convergence. 

To explore the dynamic stability and potential phase transition between the 1T-CrSH and 2H-CrSH phases, we conducted the Born-Oppenheimer molecular dynamics (BO-MD) simulations at $T=300$ K. These simulations were performed on a 27-atom unit cell (3x3x1 supercell of CrSH) using the same PAW GGA-PBE pseudopotentials. For the BO-MD simulations, the plane-wave energy cut-off was reduced to 50 Ry to optimize computational efficiency, and a 4×4×1 k-point grid was employed. The simulations were carried out in the canonical NVT ensemble. The total simulation time was set to 3.00 ps with a time step of 0.05 ps, allowing the capture of phase transition dynamics and thermal stability of 2H-CrSH as it transitioned to 1T-CrSH at room temperature.

In this study, phonon calculations were conducted using the finite displacement and supercell approaches within the DFT+$U$ framework. The Quantum ESPRESSO interface with PHONOPY software \cite{phonopy-phono3py-JPCM} was employed due to its efficiency in computing interatomic force constants (IFCs) via SCF calculations in conjunction with the ortho-atomic Hubbard $U$ parameter. The compatibility of this approach with supercell-based methods further justifies its use. In contrast, the DFPT approach with the ortho-atomic Hubbard $U$ projector is currently unsupported, presenting a limitation for phonon calculations under these conditions.

To analyze atomic displacements, a 3×3×1 supercell of the 1T-CrSH phase, consisting of 27 atoms, was constructed. The supercell approach enables the calculation of IFCs by systematically introducing small atomic displacements and performing SCF calculations to determine the resulting forces. These IFCs are significant inputs for subsequent phonon calculations. Additionally, the HIPHIVE software \cite{hiphive2019,hiphive2020} was used to enforce rotational invariance by applying both Huang and Born-Huang invariance conditions \cite{bornhuang1996}. This correction mitigates spurious imaginary frequencies often observed in the flexural phonon mode (ZA) around the $\Gamma$-point, ensuring accurate and physically consistent phonon dispersions in two-dimensional materials. 

\section{Results and discussion} \label{Sec:Results} 
In this section, we provide the results of the stabilities of CrSH in two different phases, i.e. 1T-CrSH and 2H-CrSH. Then, we discuss a possible phase transition from a half metallic to ferromagnetic monolayer. All the results were obtained by using the DFT+U method, except the BO-MD calculation in Figure~\ref{fig:MD}, in which only the PBE functional was implemented for the molecular dynamics simulations.

\subsection{Energetic, Thermal, and Dynamical Stabilities for FM 1T-CrSH and 2H-CrSH}

\begin{figure*}[bth!]
  \centering
 \includegraphics[width=1.0\linewidth]{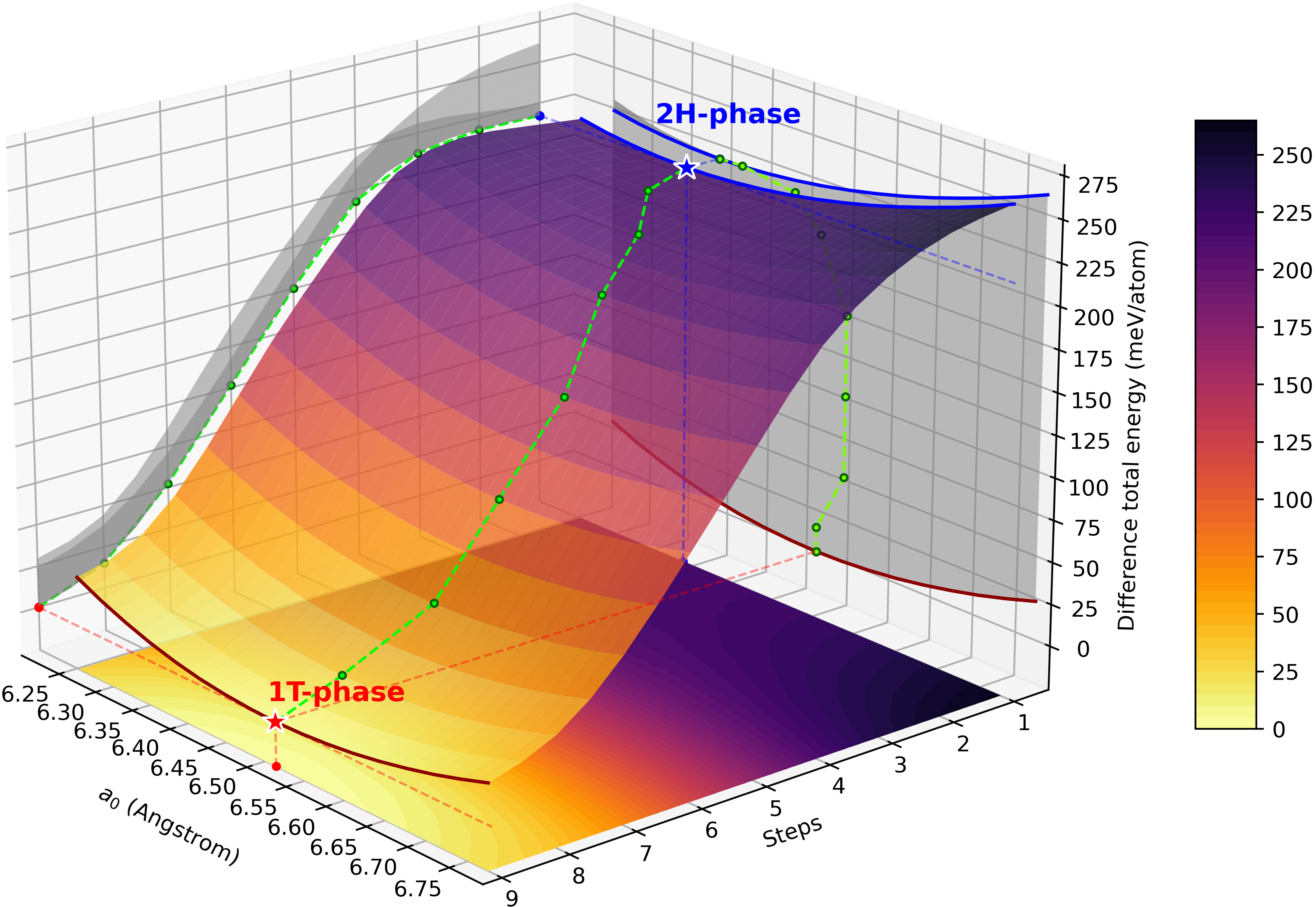}
\caption{
Relative total energy contour plot and minimum energy pathway between FM 2H-CrSH (step1) and 1T-CrSH (step9), markers by the blue and red stars, respectively.} 
\label{fig:contour}%
\end{figure*}

The structural stability and energy differences between the FM 2H-CrSH and 1T-CrSH phases are important for understanding the phase transformation mechanism. Figure~\ref{fig:contour} illustrates a 3D contour plot showing the relative total energy per atom for FM CrSH monolayers, transitioning and predicting the suitable structural transformation through a minimal energy pathway between the 2H-CrSH and 1T-CrSH phases. The X-axis represents the lattice parameter $a_0$ (equal to $b_0$), varying from 6.25 Å to 6.75 Å, which captures the contraction and expansion of the lattice during structural phase transformation. The Y-axis corresponds to nine stepwise atomic rearrangements, marking the shift from the 2H-CrSH structure (Step 1) to the 1T-CrSH structure (Step 9). Each intermediate step involves a significant change in atomic positions, particularly for the hydrogen atoms. For instance, the coordinates of Hydrogen-1 at Step 1 are (0.1667, 0.3333, 0.5519), at Step5 are (0.2500, 0.2500, 0.5476), and at Step 9 are (0.3333, 0.1667, 0.5433), indicating the gradual structural shift. 

The Z-axis represents the relative total energy, where the lowest energy is set to 0 meV/atom, corresponding to the energy minimum of FM 1T-CrSH. The color gradient indicates a scale of energy variations with lighter yellow areas marking lower energy values near 0 meV/atom, while darker purple areas denoting higher energy values, reaching up to 250 meV/atom. 
Step 1, the blue curve line represents the relative total energy for the FM 2H-CrSH, the minimum energy for FM 2H-CrSH occurs at $a_0$ =6.388 Å, marked by a blue star, which represents the equilibrium lattice parameter for this phase. The minimum energy of the 2H-CrSH phase is approximately 225 meV/atom higher than that of the 1T phase, indicating that 2H-CrSH is less stable.

As the system transitions from the 2H-CrSH phase (Step 1) to the 1T-CrSH phase (Step 9), the relative total energy progressively decreases, as shown by the green circle symbols marking the intermediate equilibrium states from Step 2 through Step 8. 
By Step 9, the red curve line represents the relative total energy for the FM 1T-CrSH structure. The minimum energy for FM 1T-CrSH is reached at $a_0$ =6.513 Å, marked by a red star, corresponding to the most stable equilibrium state. The green dashed line in Figure~\ref{fig:contour} traces the minimum energy pathway during the phase transition from 2H-CrSH to 1T-CrSH, showing a continuous decrease in energy confirms that the 1T-CrSH phase remains energetically favorable throughout the transformation. 

Compared to FM 1T-CrSH, the lattice parameters $a_0$ and $b_0$ of FM 2H-CrSH contract slightly by approximately 1.92$\%$, while the interlayer spacing between H and S layers ($d_z$) expands significantly by 13.23$\%$, from 2.298 Å to 2.602 Å. These changes in the $a_0$ and $d_z$ suggest that the displacement of hydrogen atoms significantly influences in rearranging of the Cr and S planes during the structural transformation. The structural transition also involves a shift of S-Cr-H planes from an A-B-A stacking in the 2H-CrSH phase to an A-B-C stacking in the 1T-CrSH phase. The detailed bond lengths and lattice parameters for both phases are provided in Table~\ref{Tab:a02}, where the calculated Cr-H and Cr-S bond lengths are consistent with previously reported values \cite{CrSH_1T_2024}.

\begin{table*}[tbh!]
\centering
\caption{Lattice parameters for CrSH monolayer}
\begin{tabular}{r|rrcc|crr}
\toprule
 \textbf{Lattice }  & \multicolumn{2}{c}{\textbf{FM phase}} &~& \textbf{Other work} &~& \multicolumn{2}{c}{\textbf{AFM phase}} \\
 \textbf{parameter(Å)} & \textbf{1T}   & \textbf{2H}    &~& \textbf{1T} $^{Ref.}$\cite{CrSH_1T_2024} &~& \textbf{1T}         & \textbf{2H}      \\

\midrule
a$_0$ & 6.513      & 6.388  &~& 6.44 &~   & 6.489      & 6.352     \\
c$_0$ & 20.000     & 20.000 &~& &~   & 20.000     & 20.000    \\
d$_z$ & 2.298      & 2.602  &~& &~   & 2.314      & 2.631     \\


Cr-H  & 2.070      & 2.116  &~& 2.04 &~   & 2.065      & 2.098     \\
Cr-S  & 2.363      & 2.417  &~& 2.32 &~   & 2.380      & 2.461     \\
S-H   & 2.969      & 2.602  &~& &~   & 2.981      & 2.631  \\

\bottomrule
\end{tabular}
\label{Tab:a02} 
\end{table*}

\begin{figure*}[tbh!]
  \centering
 \includegraphics[width=1.0\linewidth]{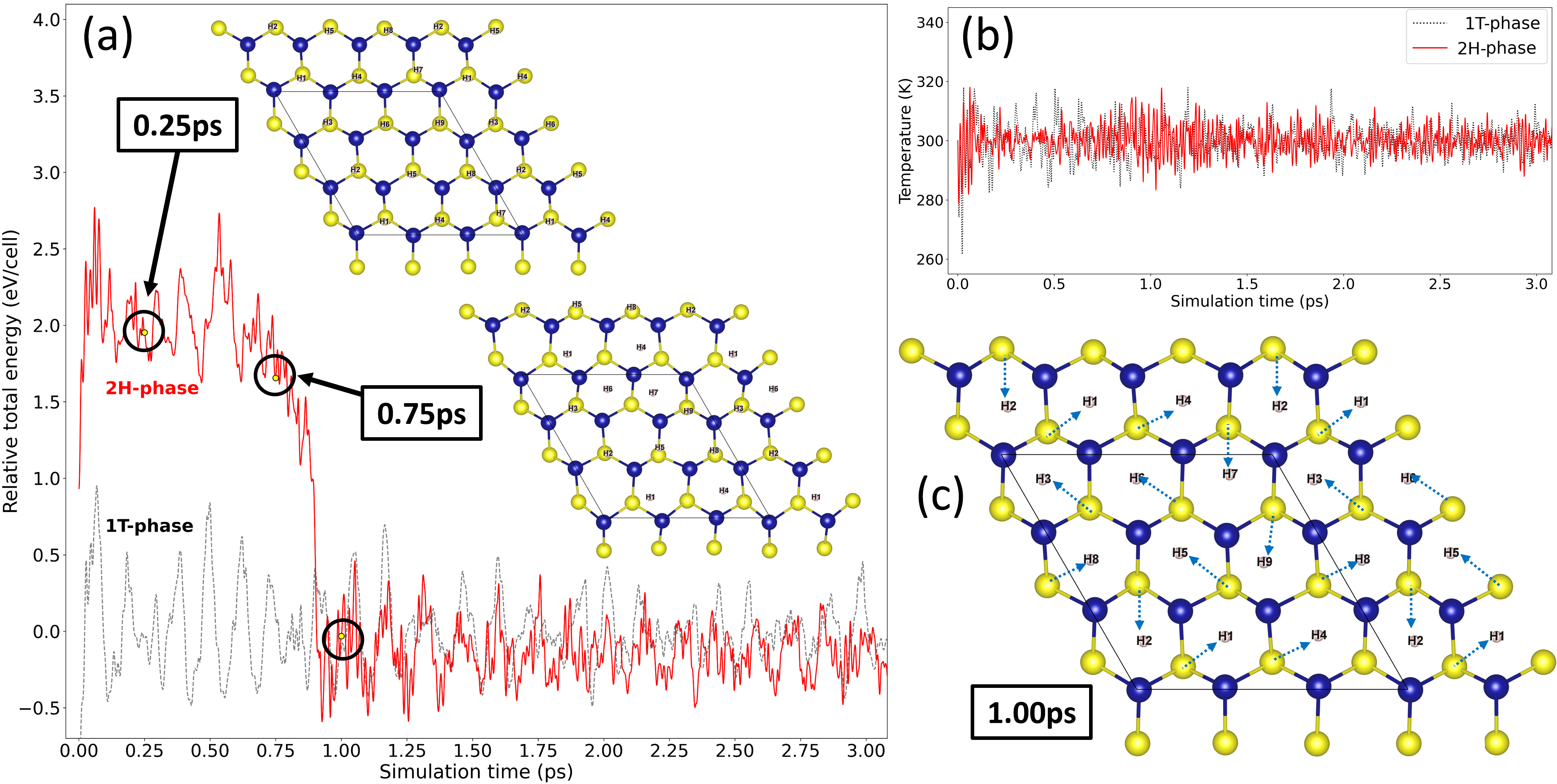}

\caption{
(a) Relative total energy using BO-MD simulation with an NVT ensemble at T = 300 K for the FM 1T-CrSH (grey dashed line) and 2H-CrSH (red line). (b) Calculated temperature profile from the BO-MD simulation. (c) Evolution of the 2H-CrSH structure after 1.0 ps of simulation, showing a complete transformation to the 1T-CrSH phase, with blue dashed arrows indicating the movement of H atoms.} 
\label{fig:MD}%
\end{figure*}

Figure~\ref{fig:MD} provides insight into the thermal stability and phase transition dynamics of FM CrSH from the metastable 2H-CrSH phase to the stable 1T-CrSH phase at 300 K using BO-MD simulations. In panel (a), illustrates the relative total energy profiles for the FM 1T-CrSH (grey dashed line) and 2H-CrSH (red line) phases as functions of simulation time. Initially, the 2H-CrSH phase has significantly higher energy compared to the 1T-CrSH phase, indicating a metastable configuration. During the simulation, the energy of the 2H-CrSH phase fluctuates but shows a significant and progressive decrease, indicating structural rearrangements that ultimately favor the 1T-CrSH configuration. By approximately 1.0 ps, the total energy of the 2H-CrSH phase aligns closely with that of the 1T-CrSH phase, suggesting a complete transformation to the energetically favorable 1T-CrSH structure. Panel (b) shows the temperature profile during the BO-MD simulation, confirming stable thermal conditions around the target temperature of 300 K. This stable temperature indicates that the phase transition observed is driven by structural instability in the 2H-CrSH phase rather than by temperature fluctuations. Panel (c) provides a snapshot of the atomic structure of CrSH at 1.0 ps, where the initial 2H-CrSH structure has fully transformed into the 1T-CrSH phase. Blue dashed arrows highlight the movement of H atoms during the transition, indicating a cooperative reconfiguration of Cr, S, and H atoms, confirming that the phase transformation occurs as a thermodynamically favorable rearrangement.

\begin{figure*}[tbh!]
  \centering
 \includegraphics[width=1.0\linewidth]{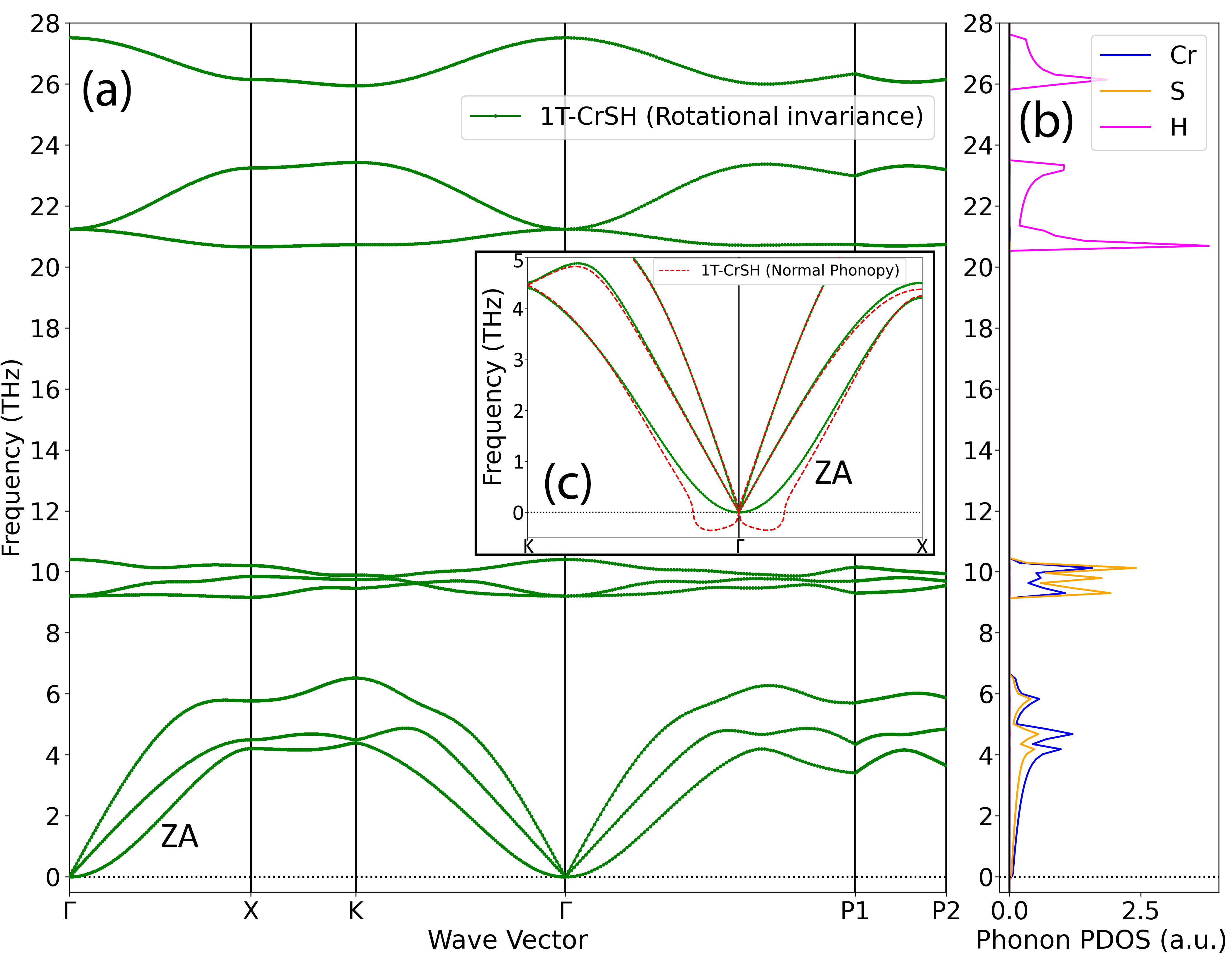}
\caption{
(a) Phonon dispersion and (b) Phonon PDOS for FM 1T-CrSH with rotational invariance corrections. (c) Phonon dispersion without rotational sum rules, showing spurious imaginary frequencies in the ZA mode near the $\Gamma$-point.} 
\label{fig:phonon}%
\end{figure*}

In Figure~\ref{fig:phonon}, we present (a) the phonon dispersion and (b) the partial phonon density of states (Phonon PDOS) for the FM 1T-CrSH phase. These calculations were performed using the finite displacement method and supercell approach with the DFT+$U$ framework via the PHONOPY software \cite{phonopy-phono3py-JPCM}, with rotational sum rules applied to the IFCs \cite{hiphive2019,hiphive2020,bornhuang1996}. The enforcement of these invariance conditions ensures the conservation of both total crystal and angular momenta, which is essential for accurate phonon behavior in 2D systems. Without the correcting rotational invariance, the ZA mode often exhibits spurious imaginary frequencies and incorrect quadratic behavior near the $\Gamma$-point, as illustrated by the red dashed line in Figure~\ref{fig:phonon}(c). This highlights the necessity of enforcing invariance conditions to correct such artifacts effectively. 
For an infinite crystal, the Huang and Born-Huang invariance conditions \cite{hiphive2019,hiphive2020,bornhuang1996} establish the acoustic sum rules required to satisfy both translational and rotational invariance of the IFCs. At equilibrium, translational invariance ensures that the total force acting on each atom vanishes, expressed as
\begin{equation}
\sum_{i} \Phi^{\alpha\beta}_{ij}=0,
\end{equation}
where $\Phi^{\alpha\beta}_{ij}$ are the second-order of IFCs between atoms $i$ and $j$ in the Cartesian direction $\alpha$ and $\beta$. 
These $\Phi^{\alpha\beta}_{ij}$ represent the second derivatives of the total potential energy of the crystal with respect to atomic displacements $u$,
\begin{equation}
\Phi^{\alpha\beta}_{ij} = \frac{\partial^2 E}{\partial^2 u^{\alpha}_{i}u^{\beta}_{j}}.
\end{equation}
This condition reflects that the total force acting on any atom must be zero due to the equilibrium configuration of all other atoms.

The rotational invariance condition introduces an additional acoustic sum rule to ensure that the lattice remains invariant under rigid rotations. For second-order IFCs, the Huang rotational invariance \cite{hiphive2019,hiphive2020,bornhuang1996} is given by
\begin{equation}
\sum_{j} \Phi^{\alpha\beta}_{ij} r^{\gamma}_{j} = \sum_{j} \Phi^{\alpha\gamma}_{ij} r^{\beta}_{j},
\end{equation}
where $r^{\beta}_{j}$ and $r^{\gamma}_{j}$ are the components of the equilibrium position vector of atom $j$ in the Cartesian directions $\beta$ and $\gamma$.

The Born-Huang sum rules \cite{hiphive2019,hiphive2020,bornhuang1996} extend the Huang invariance conditions to include higher-order symmetries associated with stress tensors and anharmonic effects
\begin{equation}
\sum_{ij} \Phi^{\alpha\beta}_{ij} r^{\gamma}_{ij} r^{\delta}_{ij} = \sum_{ij} \Phi^{\gamma\delta}_{ij} r^{\alpha}_{ij} r^{\beta}_{ij}.
\end{equation}
At equilibrium, the vanishing of the stress tensor ensures that the Huang conditions are satisfied as a subset of these rules.
The ridge regression was applied to impose these invariance conditions on the IFCs. The deviation $d=M \dot \Phi$ is minimized by solving the optimization problem
\begin{equation}
\min_{\Phi} ||M \cdot \Phi-d||^2 - \alpha||\Phi||^2,
\end{equation}
where $M$ is the constraint matrix representing the sum rules, $\Phi$ is the vector of IFC parameters, and $\alpha$ balances the magnitude of corrections and numerical stability. 

The enforcement of these invariance conditions eliminates spurious forces and corrects artifacts such as unphysical ZA-mode behavior near the $\Gamma$-point. These corrections ensure the physical accuracy of phonon dispersion calculations. These improvements are essential for accurately describing the vibrational properties of 2D materials \cite{hiphive2019,hiphive2020}.

\subsection{Electronic properties for FM 1T-CrSH and 2H-CrSH}
\begin{figure*}[tbh!]
  \centering
 \includegraphics[width=1.0\linewidth]{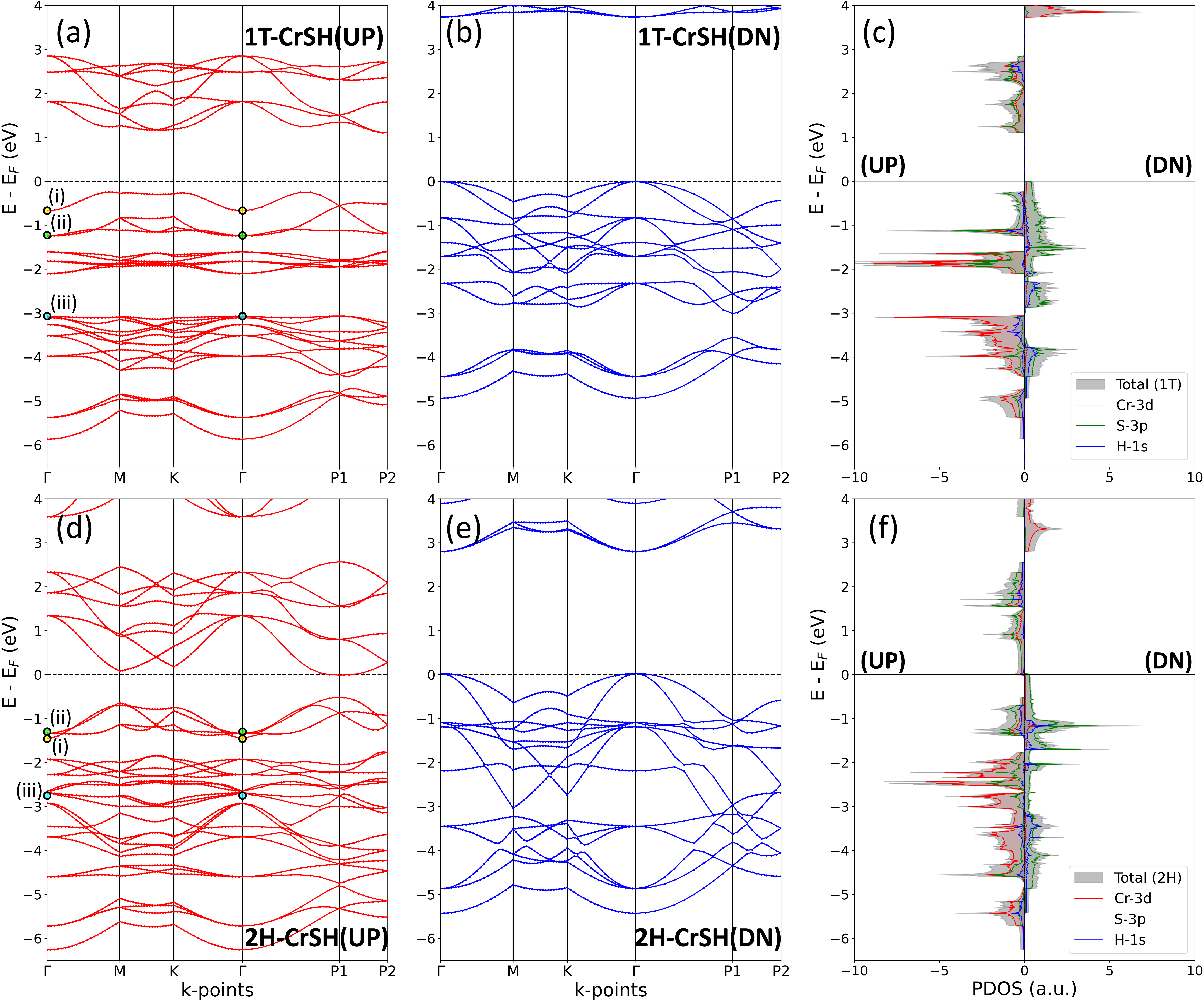}
\caption{Electronic band structures and their PDOS plots for (a)-(c) 1T-CrSH and (d)-(f) 2H-CrSH.} 
\label{fig:Band_PDOS}%
\end{figure*}

Figure~\ref{fig:Band_PDOS} illustrates the electronic properties of FM 1T-CrSH and 2H-CrSH, displaying the spin-polarized band structures for each phase. Figures~\ref{fig:Band_PDOS}(a) and~\ref{fig:Band_PDOS}(b) display the spin-up (UP) and spin-down (DN) band structures for 1T-CrSH, while Figures~\ref{fig:Band_PDOS}(d) and~\ref{fig:Band_PDOS}(e) display the spin-up and spin-down band structures for 2H-CrSH. The chosen symmetry k-point path follows: $\Gamma$ (0,0,0) $\rightarrow$ $M$(0.5,0,0) $\rightarrow$ $K$(0.3333,0.3333,0) $\rightarrow$ $\Gamma$ (0,0,0) $\rightarrow$ $P1$ (0.3333,0.5773,0) $\rightarrow$ $P2$ (0.5,0.2887,0). $P1$ and $P2$ are not high-symmetry points but were selected because they correspond to regions near the Fermi surface where the band gap closes in the 2H-CrSH phase, indicating critical points for electronic transitions. The Fermi energy level ($E_F$) is set at 0 eV, marked by the horizontal black dashed line. From 0 to -6.5 eV below the $E_F$, the spin-up channel has 28 occupied energy bands and the spin-down channel has 16, for four formula units of CrSH. This configuration gives a total magnetization of 12.0 $\mu_B$ per simulation cell, or 3.0 $\mu_B$ per Cr atom, indicating strong spin polarization in FM-CrSH.

The projected density of states (PDOS) for 1T-CrSH and 2H-CrSH are displayed in Figures~\ref{fig:Band_PDOS}(c) and~\ref{fig:Band_PDOS}(f), respectively. These PDOS plots highlight the main contributions of Cr(3d) (red), S(3p) (green), and H(1s) (blue) orbital states, with spin-up and spin-down channels represented on the left and right panels, respectively. In both phases, the PDOS provides insight into the orbital hybridizations and bonding interactions that contribute to the distinct electronic and magnetic properties observed in each phase. 

\begin{figure*}[tbh!]
  \centering
 \includegraphics[width=1.0\linewidth]{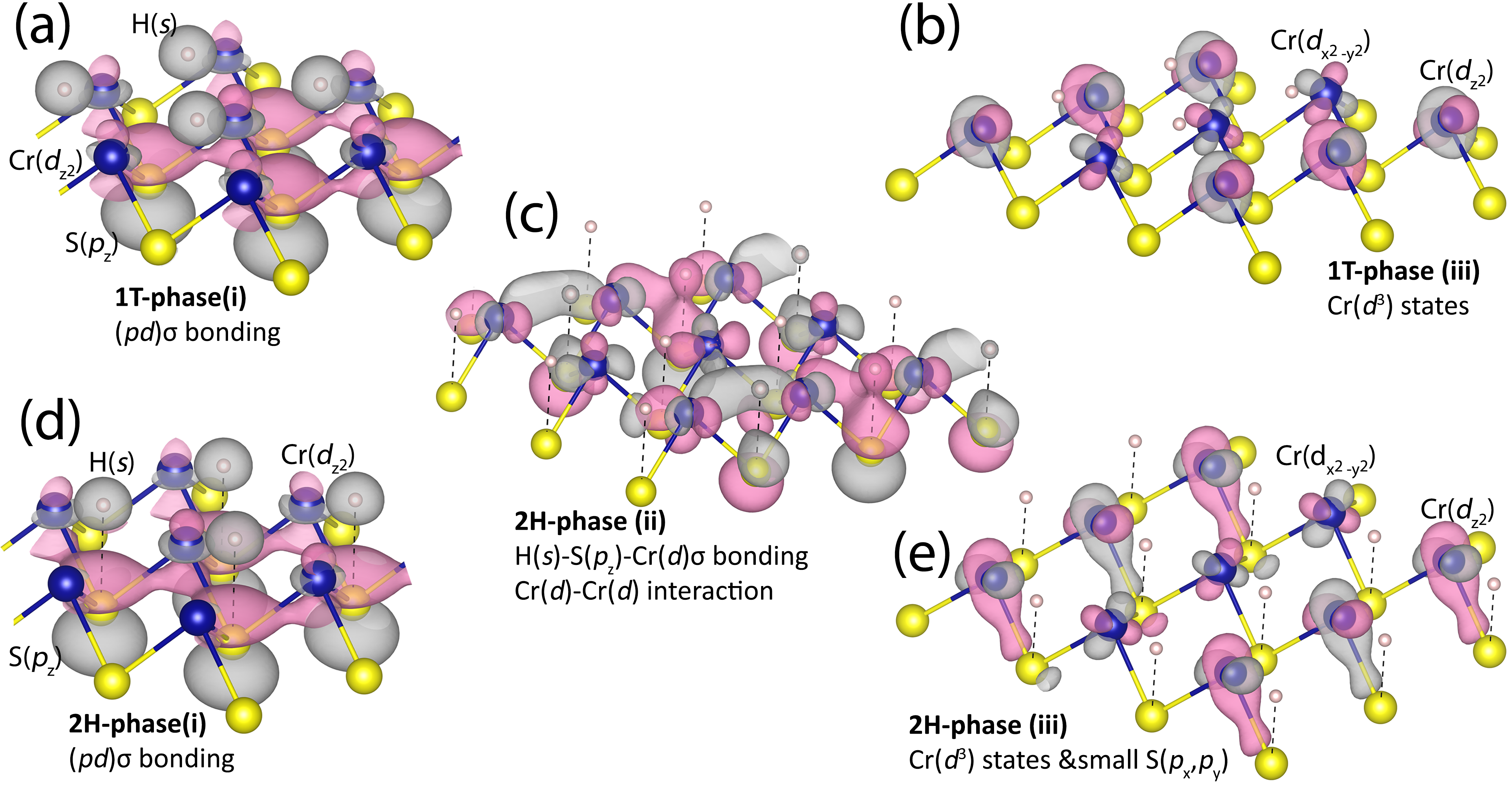}
\caption{
Molecular orbitals (MO) for selected energy states in CrSH. 
(a) 1T, band (i): Cr($d_{z^2}$) and S($p_{z}$) orbital with ($p$-$d$)$\sigma$ bonding. 
(b) 1T, band (iii): Cr($d^3$) non-bonding states. 
(c) 2H, band (ii): H($s$)-S($p_{z}$)-Cr($d$)$\sigma$ bonding and Cr($d$)-Cr($d$) interaction. 
(d) 2H, band (i): Cr($d_{z^2}$) and S($p_{z}$) orbital with ($p$-$d$)$\sigma$ bonding. 
(e) 2H, band (iii): Cr($d^3$) states with minor S($p_x$, $p_y$) contribution.}
\label{fig:MO}%
\end{figure*}

To more understating of the band structure and PDOS results, the molecular orbital (MO) theory is applied to visualize important energy states in the spin-up channels of 1T-CrSH and 2H-CrSH phases. In the 1T-CrSH phase, band(i) represents the Valence band maximum (VBM), band (ii) is the next lower-energy state, and band (iii) is a non-bonding Cr($d$) state. In the 2H-CrSH phase, the band (ii) shifts up closer to the $E_F$ and becomes the VBM, while the band (i) shifts to a lower-energy state. These selected band states are illustrated in detail with the MO shapes in Figure~\ref{fig:MO}. 

In the spin-up channel of the 1T-CrSH phase, as shown in Figures~\ref{fig:Band_PDOS}(a) and~\ref{fig:Band_PDOS}(c), the occupied states begin with contributions from Cr($dz^2$) and S($p_{z}$) orbitals, forming a ($p$-$d$)$\sigma$ bonding, and Cr($dz^2$) and H($s$) orbitals, forming a (s-d)$\sigma$* bonding, in the energy range of -0.24 to -1.24 eV, corresponding to bands (i) and (ii) in Figure~\ref{fig:Band_PDOS}(a). The band (i) is visualized by the MO shapes, shown in Figure~\ref{fig:MO}(a). Lower energy bands from -1.60 to -2.10 eV, are dominated by the S($p$) and Cr($d$) bonding states. Further down, from -3.06 to -4.43 eV, these are non-bonding Cr($d^3$) states, totaling 12 bands for four Cr atoms in the simulation cell, corresponding to the band (iii) in Figure~\ref{fig:Band_PDOS}(a) and visualized in Figure~\ref{fig:MO}(b). This strong Cr($d^3$) non-bonding interaction in the spin-up channel aligns with the high magnetic ordering, contributing significantly to the magnetic moment of 3.0 $\mu_B$ per Cr atom. Lastly, in the range of -4.72 to -5.87 eV, there is a complex hybridization involving Cr($d$), S($p$), and H($s$) orbital states.

In the spin-down channel of the 1T-CrSH phase, as shown in Figures~\ref{fig:Band_PDOS}(b) and~\ref{fig:Band_PDOS}(c), the majority of occupied states are primarily associated with S($p$) orbital states in the range of 0.00 to -2.98 eV, with minor contributions from Cr($d$) and H($s$) orbital states. Further down, from -3.56 to -4.94 eV, contributions from Cr($d$), S($p$), and H($s$) orbital states are observed, following a similar pattern to the spin-up channel but with a slight energy shift up of +0.49 eV.

The conduction band minimum (CBM) in the unoccupied states of the 1T-CrSH phase, is separated from the VBM by 1.34 eV and 3.73 eV in the spin-up and spin-down channels, respectively. This energy gap clearly indicates that the FM 1T-CrSH phase is a semiconductor, with a strong spin-polarized band gap.

The transition from the 1T-CrSH to the 2H-CrSH phase introduces significant modifications in the spin-polarized band structure and PDOS, as illustrated in Figures~\ref{fig:Band_PDOS}(d) to~\ref{fig:Band_PDOS}(f). In Figures~\ref{fig:Band_PDOS}(d) and~\ref{fig:Band_PDOS}(f), the spin-up energy bands of the 2H-CrSH phase shift to lower energies due to structural changes, specifically the reduction in the H-S distance along the z-axis from 2.969 Å to 2.602 Å and a slight decrease in the Cr-Cr distance from 3.257 Å to 3.194 Å. The CBM state, dominated by antibonding interactions between S($p_x$, $p_y$) and Cr($d_{xz}$, $d_{yz}$) orbitals, crosses $E_F$ at the $P1$ k-point, establishing a HM state. In the occupied region, the energy range from -0.52 to -1.45 eV shows a mix of ($p$-$d$)$\sigma$ and (s-d)$\sigma$* bonding interactions. The VBM band in this phase, represented by band (ii), has shifted higher in energy relative to band (i), a reversal from the 1T phase. Band (ii) is largely characterized by H($s$)-S($p_{z}$)-Cr($d$)$\sigma$ bonding interactions, which intensify due to the shorter H-S distance, as shown in the MO shapes in Figure~\ref{fig:MO}(c).

The MO shapes for band (i), illustrated in Figure~\ref{fig:MO}(d), consist of Cr($d_{z^2}$) and S($p_{z}$) orbitals, forming a ($p$-$d$)$\sigma$ bonding similar to the 1T phase but at a lower energy. The deeper energy range from -1.76 to -2.65 eV exhibits bonding interactions between S($p$) and Cr($d$) states. The Cr($d^3$) states with minor contributions from S($p_x$, $p_y$) orbitals appear in the energy range of -2.68 to -4.72 eV, corresponding to the band (iii) in Figure~\ref{fig:Band_PDOS}(d) and illustrated in Figure~\ref{fig:MO}(e). At even lower energies, between -4.83 and -6.26 eV, complex hybridization among Cr($d$), S($p$), and H($s$) orbitals is observed. 

In the spin-down channel of 2H-CrSH, as displayed in Figures~\ref{fig:Band_PDOS}(e) and~\ref{fig:Band_PDOS}(f), the occupied states primarily arise from S($p$) and H($s$) orbitals, with minor contributions from Cr($d$) orbitals. This bonding interaction spans a broad energy range from 0.0 to -5.43 eV, creating a more extensive band distribution compared to the 1T-CrSH phase. The widened bands in the spin-down channel of the 2H-CrSH phase indicate enhanced orbital overlap and hybridization, which may be related to the metallic character seen in the spin-up channel. This broader band structure in the spin-down channel highlights the complex nature of electron delocalization in the 2H-CrSH phase, further confirming its HM state and suggesting potential utility in spintronic applications.


\begin{figure*}[tbh!]
  \centering
 \includegraphics[width=0.85\linewidth]{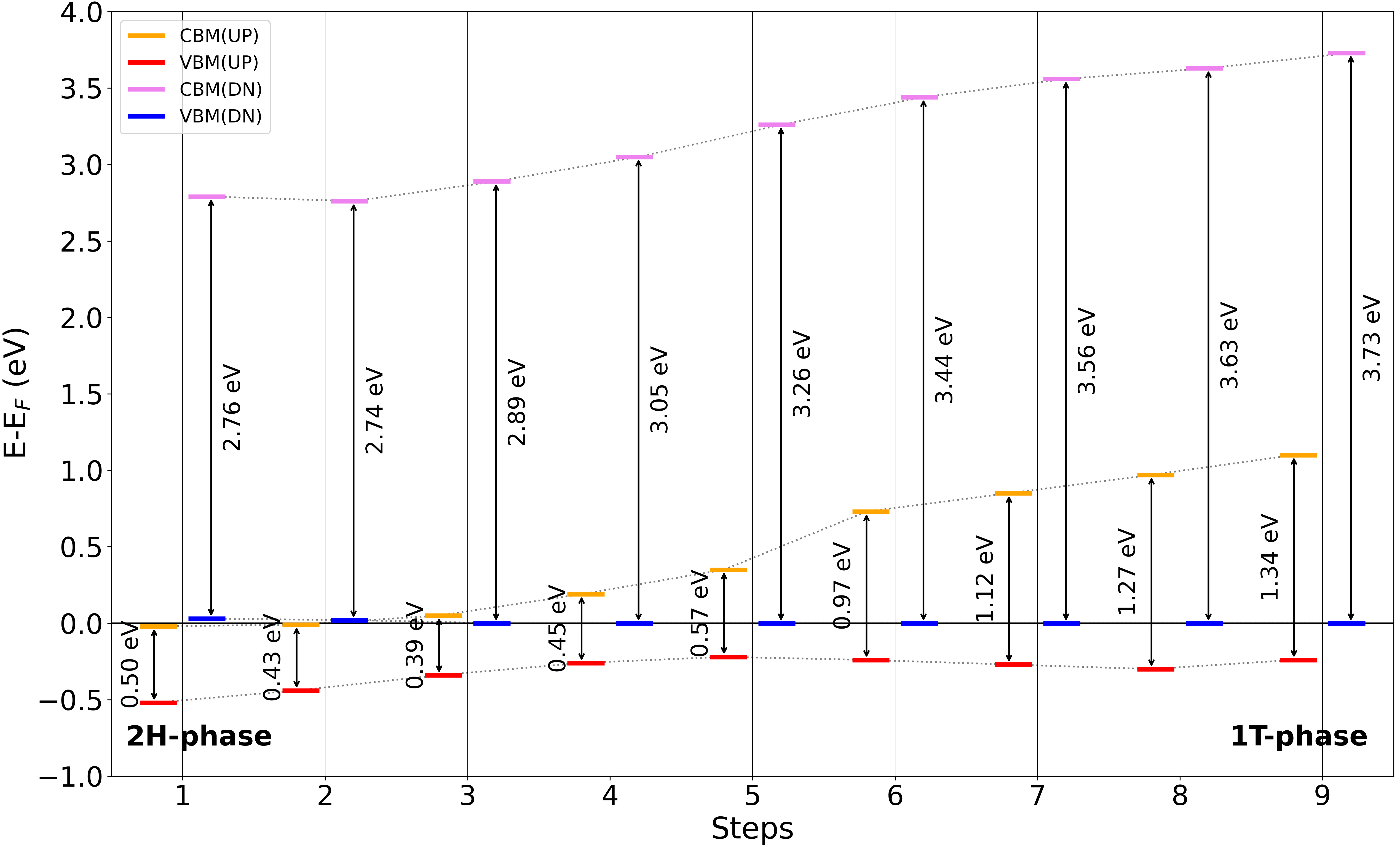}
\caption{
Evolution of the CBM and VBM for CrSH during the transition from 2H-CrSH (Step1) to 1T-CrSH (Step9). The spin-up and spin-down contributions are shown separately, with CBM (UP) in orange, VBM (UP) in red, CBM (DN) in purple, and VBM (DN) in blue.} 
\label{fig:Eg}%
\end{figure*}

Figure~\ref{fig:Eg} summarizes the evolution of the CBM and VBM for CrSH during the phase transition from the 2H-CrSH phase (Step1) to the 1T-CrSH phase (Step9). We will refer to this phase transition as the FM half-metallic to FM semiconductor phase transition, accompanying the structural phase transition from 2H to 1T phases. This transition reveals significant changes in the band gap and spin polarization between these phases, emphasizing the influence of structural modifications on the electronic properties of FM CrSH monolayer. 

In the 2H-CrSH phase (Step 1), the CBM in the spin-up channel crosses the $E_F$, resulting in a metallic behavior. Meanwhile, the spin-down CBM-VBM separation is 2.76 eV, indicating that the 2H-CrSH exhibits an HM phase. This asymmetry between spin-up and spin-down channels is a hallmark of half-metallicity, making the 2H-CrSH phase suitable for spin-polarized conduction. 
As the transition progresses from Step 1 to Step 9, the CBM states in both spin-up and spin-down channels exhibit a noticeable upward shift, resulting in a gradual increase of the band gap. This evolution reflects the material's transformation toward semiconducting behavior. 

By the 1T-CrSH phase (Step 9), the CBM-VBM separation becomes 1.34 eV in the spin-up channel and 3.73 eV in the spin-down channel, indicating a wide band gap in both spin channels. The transition from HM behavior in 2H-CrSH to semiconducting behavior in 1T-CrSH underscores the strong coupling between structural and electronic properties in FM CrSH monolayer. 
The asymmetry between spin-up and spin-down channels throughout the intermediate steps (Step 2 to Step 8) further suggests potential for spintronic applications. The tunable electronic properties during this transition provide an opportunity to design devices capable of selectively controlling spin polarization and conduction properties. 

The unstable 2H-CrSH phase exhibits an HM behavior due to the closure of the spin-up band gap, which could enable spin-polarized conduction. As the system transitions to the stable 1T-CrSH leads to distinct shifts in CBM and VBM states, demonstrating the impact of structural changes on the electronic band structure and emphasizing the material’s promise for spintronic device applications. This tunable electronic behavior in FM CrSH, controlled by phase transition, highlights its potential in applications requiring selective spin polarization and adjustable electronic properties.

\section{Conclusion} \label{Sec:Conclusion}
This study provides a comprehensive analysis of the structural, electronic, magnetic, and vibrational properties of the FM CrSH monolayer in its 1T and 2H phases using \textit{ab}-initio first-principles DFT+$U$ calculations. The results highlight the superior stability of the FM 1T-CrSH phase, which exhibits semiconducting behavior with a band gap of 1.1 eV. In contrast, the FM 2H-CrSH phase is identified as metastable, transitioning into an HM state during the phase transformation. The BO-MD simulations at 300 K confirm that the 2H-CrSH phase is dynamically unstable, undergoing a rapid phase transition to the stable 1T-CrSH phase. 

The phonon dispersion analysis, performed using the finite displacement method, with corrections for Huang and Born-Huang rotational invariance conditions, resolves artifacts such as imaginary frequencies in the flexural ZA mode near the $\Gamma$-point. These corrections ensure physically consistent vibrational properties, critical for the accurate modeling of 2D material systems. 

The electronic structure analysis highlights the strong spin polarization of the FM 1T-CrSH phase, which exhibits a magnetic moment of 3.0 $\mu_B$ per Cr atom, making it highly suitable for spintronic applications. The phase transition from 2H-CrSH to 1T-CrSH introduces significant electronic modifications, including the emergence of half-metallicity in the 2H phase, further emphasizing its potential utility in spintronic devices.

These findings establish the FM 1T-CrSH monolayer as a promising candidate for next-generation spintronic and valleytronic technologies. Moreover, the tunable electronic properties enabled by phase transitions further enhance its applicability in advanced technologies. Future experimental studies are encouraged to validate these theoretical predictions and explore practical implementations of CrSH-based devices. 

\section*{Acknowledgment}
This research was supported by the Ratchadapisek Somphot Fund for Postdoctoral Fellowship, and the Second Century Fund (C2F) at Chulalongkorn University. We gratefully acknowledge the computing infrastructure provided by the National Science and Technology Development Agency (NSTDA), Chulalongkorn University (CU), and CUAASC. Additional support was provided by the National Science Research and Innovation Fund (NSRF) through PMUC [B05F650021, B37G660013], Thailand. Access to the computational facilities was facilitated via the e-Science platform (URL: www.e-science.in.th).

\bibliography{cas-refs}

\providecommand{\latin}[1]{#1}
\makeatletter
\providecommand{\doi}
  {\begingroup\let\do\@makeother\dospecials
  \catcode`\{=1 \catcode`\}=2 \doi@aux}
\providecommand{\doi@aux}[1]{\endgroup\texttt{#1}}
\makeatother
\providecommand*\mcitethebibliography{\thebibliography}
\csname @ifundefined\endcsname{endmcitethebibliography}  {\let\endmcitethebibliography\endthebibliography}{}
\begin{mcitethebibliography}{64}
\providecommand*\natexlab[1]{#1}
\providecommand*\mciteSetBstSublistMode[1]{}
\providecommand*\mciteSetBstMaxWidthForm[2]{}
\providecommand*\mciteBstWouldAddEndPuncttrue
  {\def\EndOfBibitem{\unskip.}}
\providecommand*\mciteBstWouldAddEndPunctfalse
  {\let\EndOfBibitem\relax}
\providecommand*\mciteSetBstMidEndSepPunct[3]{}
\providecommand*\mciteSetBstSublistLabelBeginEnd[3]{}
\providecommand*\EndOfBibitem{}
\mciteSetBstSublistMode{f}
\mciteSetBstMaxWidthForm{subitem}{(\alph{mcitesubitemcount})}
\mciteSetBstSublistLabelBeginEnd
  {\mcitemaxwidthsubitemform\space}
  {\relax}
  {\relax}

\bibitem[Gupta \latin{et~al.}(2015)Gupta, Sakthivel, and Seal]{nanoelect_2015}
Gupta,~A.; Sakthivel,~T.; Seal,~S. Recent development in 2D materials beyond graphene. \emph{Progress in Materials Science} \textbf{2015}, \emph{73}, 44--126\relax
\mciteBstWouldAddEndPuncttrue
\mciteSetBstMidEndSepPunct{\mcitedefaultmidpunct}
{\mcitedefaultendpunct}{\mcitedefaultseppunct}\relax
\EndOfBibitem
\bibitem[B{\"o}ker \latin{et~al.}(2001)B{\"o}ker, Severin, M{\"u}ller, Janowitz, Manzke, Vo{\ss}, Kr{\"u}ger, Mazur, and Pollmann]{MoS2_2001}
B{\"o}ker,~T.; Severin,~R.; M{\"u}ller,~A.; Janowitz,~C.; Manzke,~R.; Vo{\ss},~D.; Kr{\"u}ger,~P.; Mazur,~A.; Pollmann,~J. Band structure of MoS2, MoSe2, and $\alpha$- MoTe2: Angle-resolved photoelectron spectroscopy and ab initio calculations. \emph{Physical Review B} \textbf{2001}, \emph{64}, 235305\relax
\mciteBstWouldAddEndPuncttrue
\mciteSetBstMidEndSepPunct{\mcitedefaultmidpunct}
{\mcitedefaultendpunct}{\mcitedefaultseppunct}\relax
\EndOfBibitem
\bibitem[Singh \latin{et~al.}(2019)Singh, Singh, Kim, Yeom, and Nalwa]{MoS2_Optoelec2019}
Singh,~E.; Singh,~P.; Kim,~K.~S.; Yeom,~G.~Y.; Nalwa,~H.~S. Flexible Molybdenum Disulfide (MoS2) Atomic Layers for Wearable Electronics and Optoelectronics. \emph{ACS Applied Materials \& Interfaces} \textbf{2019}, \emph{11}, 11061--11105\relax
\mciteBstWouldAddEndPuncttrue
\mciteSetBstMidEndSepPunct{\mcitedefaultmidpunct}
{\mcitedefaultendpunct}{\mcitedefaultseppunct}\relax
\EndOfBibitem
\bibitem[Pham \latin{et~al.}(2022)Pham, Bodepudi, Shehzad, Liu, Xu, Yu, and Duan]{Optoelectronics2022}
Pham,~P.~V.; Bodepudi,~S.~C.; Shehzad,~K.; Liu,~Y.; Xu,~Y.; Yu,~B.; Duan,~X. 2D Heterostructures for Ubiquitous Electronics and Optoelectronics: Principles, Opportunities, and Challenges. \emph{Chemical Reviews} \textbf{2022}, \emph{122}, 6514--6613, PMID: 35133801\relax
\mciteBstWouldAddEndPuncttrue
\mciteSetBstMidEndSepPunct{\mcitedefaultmidpunct}
{\mcitedefaultendpunct}{\mcitedefaultseppunct}\relax
\EndOfBibitem
\bibitem[Liu \latin{et~al.}(2020)Liu, Zhang, Ma, Tong, Han, and Wang]{storage2020}
Liu,~X.; Zhang,~X.; Ma,~S.; Tong,~S.; Han,~X.; Wang,~H. Flexible amorphous MoS2 nanoflakes/N-doped carbon microtubes/reduced graphite oxide composite paper as binder free anode for full cell lithium ion batteries. \emph{Electrochimica Acta} \textbf{2020}, \emph{333}, 135568\relax
\mciteBstWouldAddEndPuncttrue
\mciteSetBstMidEndSepPunct{\mcitedefaultmidpunct}
{\mcitedefaultendpunct}{\mcitedefaultseppunct}\relax
\EndOfBibitem
\bibitem[Jiang \latin{et~al.}(2022)Jiang, Liu, Xiao, Qian, Sun, Zeng, and Wang]{storage2021}
Jiang,~D.; Liu,~Z.; Xiao,~Z.; Qian,~Z.; Sun,~Y.; Zeng,~Z.; Wang,~R. Flexible electronics based on 2D transition metal dichalcogenides. \emph{J. Mater. Chem. A} \textbf{2022}, \emph{10}, 89--121\relax
\mciteBstWouldAddEndPuncttrue
\mciteSetBstMidEndSepPunct{\mcitedefaultmidpunct}
{\mcitedefaultendpunct}{\mcitedefaultseppunct}\relax
\EndOfBibitem
\bibitem[Ho \latin{et~al.}(2004)Ho, Yu, Lin, Yu, and Li]{WS2MoS2_2004}
Ho,~W.; Yu,~J.~C.; Lin,~J.; Yu,~J.; Li,~P. Preparation and photocatalytic behavior of MoS2 and WS2 nanocluster sensitized TiO2. \emph{Langmuir} \textbf{2004}, \emph{20}, 5865--5869\relax
\mciteBstWouldAddEndPuncttrue
\mciteSetBstMidEndSepPunct{\mcitedefaultmidpunct}
{\mcitedefaultendpunct}{\mcitedefaultseppunct}\relax
\EndOfBibitem
\bibitem[Chandrasekaran \latin{et~al.}(2019)Chandrasekaran, Yao, Deng, Bowen, Zhang, Chen, Lin, Peng, and Zhang]{MoS2_App2019}
Chandrasekaran,~S.; Yao,~L.; Deng,~L.; Bowen,~C.; Zhang,~Y.; Chen,~S.; Lin,~Z.; Peng,~F.; Zhang,~P. Recent advances in metal sulfides: from controlled fabrication to electrocatalytic{,} photocatalytic and photoelectrochemical water splitting and beyond. \emph{Chem. Soc. Rev.} \textbf{2019}, \emph{48}, 4178--4280\relax
\mciteBstWouldAddEndPuncttrue
\mciteSetBstMidEndSepPunct{\mcitedefaultmidpunct}
{\mcitedefaultendpunct}{\mcitedefaultseppunct}\relax
\EndOfBibitem
\bibitem[Mahatha \latin{et~al.}(2012)Mahatha, Patel, and Menon]{MoS2-1_2012}
Mahatha,~S.~K.; Patel,~K.~D.; Menon,~K. S.~R. Electronic structure investigation of MoS2 and MoSe2 using angle-resolved photoemission spectroscopy and ab initio band structure studies. \emph{Journal of Physics: Condensed Matter} \textbf{2012}, \emph{24}, 475504\relax
\mciteBstWouldAddEndPuncttrue
\mciteSetBstMidEndSepPunct{\mcitedefaultmidpunct}
{\mcitedefaultendpunct}{\mcitedefaultseppunct}\relax
\EndOfBibitem
\bibitem[Stephenson \latin{et~al.}(2014)Stephenson, Li, Olsen, and Mitlin]{MoS2_2014}
Stephenson,~T.; Li,~Z.; Olsen,~B.; Mitlin,~D. Lithium ion battery applications of molybdenum disulfide (MoS 2) nanocomposites. \emph{Energy \& Environmental Science} \textbf{2014}, \emph{7}, 209--231\relax
\mciteBstWouldAddEndPuncttrue
\mciteSetBstMidEndSepPunct{\mcitedefaultmidpunct}
{\mcitedefaultendpunct}{\mcitedefaultseppunct}\relax
\EndOfBibitem
\bibitem[Chen \latin{et~al.}(2014)Chen, Ouyang, Yuan, Li, and Wang]{MoS2_HM_2014}
Chen,~Q.; Ouyang,~Y.; Yuan,~S.; Li,~R.; Wang,~J. Uniformly wetting deposition of Co atoms on MoS2 monolayer: a promising two-dimensional robust half-metallic ferromagnet. \emph{ACS applied materials \& interfaces} \textbf{2014}, \emph{6}, 16835--16840\relax
\mciteBstWouldAddEndPuncttrue
\mciteSetBstMidEndSepPunct{\mcitedefaultmidpunct}
{\mcitedefaultendpunct}{\mcitedefaultseppunct}\relax
\EndOfBibitem
\bibitem[Wang \latin{et~al.}(2020)Wang, Shu, Zhou, Wang, and Chen]{MoSe2_2020}
Wang,~W.; Shu,~H.; Zhou,~D.; Wang,~J.; Chen,~X. Ultrafast nucleation and growth of high-quality monolayer MoSe2 crystals via vapor-liquid-solid mechanism. \emph{Nanotechnology} \textbf{2020}, \emph{31}, 335601\relax
\mciteBstWouldAddEndPuncttrue
\mciteSetBstMidEndSepPunct{\mcitedefaultmidpunct}
{\mcitedefaultendpunct}{\mcitedefaultseppunct}\relax
\EndOfBibitem
\bibitem[Sha \latin{et~al.}(2022)Sha, Maity, Rajaji, Liu, and Bhattacharyya]{MoSe2_2022}
Sha,~R.; Maity,~P.~C.; Rajaji,~U.; Liu,~T.-Y.; Bhattacharyya,~T.~K. Review—MoSe2 Nanostructures and Related Electrodes for Advanced Supercapacitor Developments. \emph{Journal of The Electrochemical Society} \textbf{2022}, \emph{169}, 013503\relax
\mciteBstWouldAddEndPuncttrue
\mciteSetBstMidEndSepPunct{\mcitedefaultmidpunct}
{\mcitedefaultendpunct}{\mcitedefaultseppunct}\relax
\EndOfBibitem
\bibitem[Tongay \latin{et~al.}(2012)Tongay, Varnoosfaderani, Appleton, Wu, and Hebard]{MoS2-2_2012}
Tongay,~S.; Varnoosfaderani,~S.~S.; Appleton,~B.~R.; Wu,~J.; Hebard,~A.~F. Magnetic properties of MoS2: Existence of ferromagnetism. \emph{Applied Physics Letters} \textbf{2012}, \emph{101}, 123105\relax
\mciteBstWouldAddEndPuncttrue
\mciteSetBstMidEndSepPunct{\mcitedefaultmidpunct}
{\mcitedefaultendpunct}{\mcitedefaultseppunct}\relax
\EndOfBibitem
\bibitem[Schaibley \latin{et~al.}(2016)Schaibley, Yu, Clark, Rivera, Ross, Seyler, Yao, and Xu]{valley2016}
Schaibley,~J.~R.; Yu,~H.; Clark,~G.; Rivera,~P.; Ross,~J.~S.; Seyler,~K.~L.; Yao,~W.; Xu,~X. Valleytronics in 2D materials. \emph{Nature Reviews Materials} \textbf{2016}, \emph{1}, 1--15\relax
\mciteBstWouldAddEndPuncttrue
\mciteSetBstMidEndSepPunct{\mcitedefaultmidpunct}
{\mcitedefaultendpunct}{\mcitedefaultseppunct}\relax
\EndOfBibitem
\bibitem[Wang and Wang(2015)Wang, and Wang]{MoS2_HM_2015}
Wang,~S.; Wang,~J. Spin and valley half-metal state in MoS2 monolayer. \emph{Physica B: Condensed Matter} \textbf{2015}, \emph{458}, 22--26\relax
\mciteBstWouldAddEndPuncttrue
\mciteSetBstMidEndSepPunct{\mcitedefaultmidpunct}
{\mcitedefaultendpunct}{\mcitedefaultseppunct}\relax
\EndOfBibitem
\bibitem[Krstajić \latin{et~al.}(2016)Krstajić, Vasilopoulos, and Tahir]{MoS2_2016}
Krstajić,~P.; Vasilopoulos,~P.; Tahir,~M. Spin- and valley-polarized transport through ferromagnetic and antiferromagnetic barriers on monolayer MoS2. \emph{Physica E: Low-dimensional Systems and Nanostructures} \textbf{2016}, \emph{75}, 317--321\relax
\mciteBstWouldAddEndPuncttrue
\mciteSetBstMidEndSepPunct{\mcitedefaultmidpunct}
{\mcitedefaultendpunct}{\mcitedefaultseppunct}\relax
\EndOfBibitem
\bibitem[Rahman \latin{et~al.}(2017)Rahman, Rahman, and García-Suárez]{MoS2_HM_2017}
Rahman,~A.~U.; Rahman,~G.; García-Suárez,~V.~M. Development of spontaneous magnetism and half-metallicity in monolayer MoS2. \emph{Journal of Magnetism and Magnetic Materials} \textbf{2017}, \emph{443}, 343--351\relax
\mciteBstWouldAddEndPuncttrue
\mciteSetBstMidEndSepPunct{\mcitedefaultmidpunct}
{\mcitedefaultendpunct}{\mcitedefaultseppunct}\relax
\EndOfBibitem
\bibitem[Jiang \latin{et~al.}(2018)Jiang, Wang, Zhang, Wang, Chen, and Wan]{MoS2_HM_2018}
Jiang,~C.; Wang,~Y.; Zhang,~Y.; Wang,~H.; Chen,~Q.; Wan,~J. Robust half-metallic magnetism in two-dimensional Fe/MoS2. \emph{The Journal of Physical Chemistry C} \textbf{2018}, \emph{122}, 21617--21622\relax
\mciteBstWouldAddEndPuncttrue
\mciteSetBstMidEndSepPunct{\mcitedefaultmidpunct}
{\mcitedefaultendpunct}{\mcitedefaultseppunct}\relax
\EndOfBibitem
\bibitem[Rahman \latin{et~al.}(2021)Rahman, Torres, Khan, and Lu]{MoS2_AFM_2021}
Rahman,~S.; Torres,~J.~F.; Khan,~A.~R.; Lu,~Y. Recent developments in van der Waals antiferromagnetic 2D materials: Synthesis, characterization, and device implementation. \emph{ACS nano} \textbf{2021}, \emph{15}, 17175--17213\relax
\mciteBstWouldAddEndPuncttrue
\mciteSetBstMidEndSepPunct{\mcitedefaultmidpunct}
{\mcitedefaultendpunct}{\mcitedefaultseppunct}\relax
\EndOfBibitem
\bibitem[Gao \latin{et~al.}(2013)Gao, Xue, Mao, Wang, Xu, and Xue]{VS2_2013}
Gao,~D.; Xue,~Q.; Mao,~X.; Wang,~W.; Xu,~Q.; Xue,~D. Ferromagnetism in ultrathin VS2 nanosheets. \emph{J. Mater. Chem. C} \textbf{2013}, \emph{1}, 5909--5916\relax
\mciteBstWouldAddEndPuncttrue
\mciteSetBstMidEndSepPunct{\mcitedefaultmidpunct}
{\mcitedefaultendpunct}{\mcitedefaultseppunct}\relax
\EndOfBibitem
\bibitem[Guo \latin{et~al.}(2017)Guo, Deng, Sun, Li, Zhao, Wu, Chu, Zhang, Pan, Zheng, Wu, Jin, Wu, and Xie]{VS2_2017}
Guo,~Y.; Deng,~H.; Sun,~X.; Li,~X.; Zhao,~J.; Wu,~J.; Chu,~W.; Zhang,~S.; Pan,~H.; Zheng,~X.; Wu,~X.; Jin,~C.; Wu,~C.; Xie,~Y. Modulation of Metal and Insulator States in 2D Ferromagnetic VS2 by van der Waals Interaction Engineering. \emph{Advanced Materials} \textbf{2017}, \emph{29}, 1700715\relax
\mciteBstWouldAddEndPuncttrue
\mciteSetBstMidEndSepPunct{\mcitedefaultmidpunct}
{\mcitedefaultendpunct}{\mcitedefaultseppunct}\relax
\EndOfBibitem
\bibitem[Feng \latin{et~al.}(2018)Feng, Wu, Han, and Gao]{CoCl2_2018}
Feng,~Y.; Wu,~X.; Han,~J.; Gao,~G. Robust half-metallicities and perfect spin transport properties in 2D transition metal dichlorides. \emph{Journal of Materials Chemistry C} \textbf{2018}, \emph{6}, 4087--4094\relax
\mciteBstWouldAddEndPuncttrue
\mciteSetBstMidEndSepPunct{\mcitedefaultmidpunct}
{\mcitedefaultendpunct}{\mcitedefaultseppunct}\relax
\EndOfBibitem
\bibitem[Lv \latin{et~al.}(2019)Lv, Lu, Luo, Zhu, and Sun]{CoBr2_2019}
Lv,~H.; Lu,~W.; Luo,~X.; Zhu,~X.; Sun,~Y. Strain-and carrier-tunable magnetic properties of a two-dimensional intrinsically ferromagnetic semiconductor: CoBr 2 monolayer. \emph{Physical Review B} \textbf{2019}, \emph{99}, 134416\relax
\mciteBstWouldAddEndPuncttrue
\mciteSetBstMidEndSepPunct{\mcitedefaultmidpunct}
{\mcitedefaultendpunct}{\mcitedefaultseppunct}\relax
\EndOfBibitem
\bibitem[Liu \latin{et~al.}(2023)Liu, Wang, Zhang, Ma, Chen, Liu, Zhang, Feng, Cheng, Zhao, \latin{et~al.} others]{CoCl2_2023}
Liu,~H.; Wang,~A.; Zhang,~P.; Ma,~C.; Chen,~C.; Liu,~Z.; Zhang,~Y.-Q.; Feng,~B.; Cheng,~P.; Zhao,~J.; others Atomic-scale manipulation of single-polaron in a two-dimensional semiconductor. \emph{Nature Communications} \textbf{2023}, \emph{14}, 3690\relax
\mciteBstWouldAddEndPuncttrue
\mciteSetBstMidEndSepPunct{\mcitedefaultmidpunct}
{\mcitedefaultendpunct}{\mcitedefaultseppunct}\relax
\EndOfBibitem
\bibitem[Kan \latin{et~al.}(2014)Kan, Adhikari, and Sun]{MnS2_2014}
Kan,~M.; Adhikari,~S.; Sun,~Q. Ferromagnetism in mnx 2 (x= s, se) monolayers. \emph{Physical Chemistry Chemical Physics} \textbf{2014}, \emph{16}, 4990--4994\relax
\mciteBstWouldAddEndPuncttrue
\mciteSetBstMidEndSepPunct{\mcitedefaultmidpunct}
{\mcitedefaultendpunct}{\mcitedefaultseppunct}\relax
\EndOfBibitem
\bibitem[Trivedi \latin{et~al.}(2020)Trivedi, Turgut, Qin, Sayyad, Hajra, Howell, Liu, Yang, Patoary, Li, \latin{et~al.} others]{Spin_2020}
Trivedi,~D.~B.; Turgut,~G.; Qin,~Y.; Sayyad,~M.~Y.; Hajra,~D.; Howell,~M.; Liu,~L.; Yang,~S.; Patoary,~N.~H.; Li,~H.; others Room-temperature synthesis of 2D Janus crystals and their heterostructures. \emph{Advanced materials} \textbf{2020}, \emph{32}, 2006320\relax
\mciteBstWouldAddEndPuncttrue
\mciteSetBstMidEndSepPunct{\mcitedefaultmidpunct}
{\mcitedefaultendpunct}{\mcitedefaultseppunct}\relax
\EndOfBibitem
\bibitem[Lu \latin{et~al.}(2017)Lu, Zhu, Xiao, Chuu, Han, Chiu, Cheng, Yang, Wei, Yang, \latin{et~al.} others]{lu2017janus}
Lu,~A.-Y.; Zhu,~H.; Xiao,~J.; Chuu,~C.-P.; Han,~Y.; Chiu,~M.-H.; Cheng,~C.-C.; Yang,~C.-W.; Wei,~K.-H.; Yang,~Y.; others Janus monolayers of transition metal dichalcogenides. \emph{Nature nanotechnology} \textbf{2017}, \emph{12}, 744--749\relax
\mciteBstWouldAddEndPuncttrue
\mciteSetBstMidEndSepPunct{\mcitedefaultmidpunct}
{\mcitedefaultendpunct}{\mcitedefaultseppunct}\relax
\EndOfBibitem
\bibitem[Sant \latin{et~al.}(2020)Sant, Gay, Marty, Lisi, Harrabi, Vergnaud, Dau, Weng, Coraux, Gauthier, \latin{et~al.} others]{sant2020synthesis}
Sant,~R.; Gay,~M.; Marty,~A.; Lisi,~S.; Harrabi,~R.; Vergnaud,~C.; Dau,~M.~T.; Weng,~X.; Coraux,~J.; Gauthier,~N.; others Synthesis of epitaxial monolayer Janus SPtSe. \emph{npj 2D Materials and Applications} \textbf{2020}, \emph{4}, 41\relax
\mciteBstWouldAddEndPuncttrue
\mciteSetBstMidEndSepPunct{\mcitedefaultmidpunct}
{\mcitedefaultendpunct}{\mcitedefaultseppunct}\relax
\EndOfBibitem
\bibitem[Zhang \latin{et~al.}(2019)Zhang, Nie, Sanvito, and Du]{VSSe_2019}
Zhang,~C.; Nie,~Y.; Sanvito,~S.; Du,~A. First-principles prediction of a room-temperature ferromagnetic Janus VSSe monolayer with piezoelectricity, ferroelasticity, and large valley polarization. \emph{Nano letters} \textbf{2019}, \emph{19}, 1366--1370\relax
\mciteBstWouldAddEndPuncttrue
\mciteSetBstMidEndSepPunct{\mcitedefaultmidpunct}
{\mcitedefaultendpunct}{\mcitedefaultseppunct}\relax
\EndOfBibitem
\bibitem[Kong \latin{et~al.}(2019)Kong, Stolze, Timmons, Tao, Ni, Guo, Yang, Prozorov, and Cava]{VI3-1_2019}
Kong,~T.; Stolze,~K.; Timmons,~E.~I.; Tao,~J.; Ni,~D.; Guo,~S.; Yang,~Z.; Prozorov,~R.; Cava,~R.~J. VI3—a new layered ferromagnetic semiconductor. \emph{Advanced Materials} \textbf{2019}, \emph{31}, 1808074\relax
\mciteBstWouldAddEndPuncttrue
\mciteSetBstMidEndSepPunct{\mcitedefaultmidpunct}
{\mcitedefaultendpunct}{\mcitedefaultseppunct}\relax
\EndOfBibitem
\bibitem[Tian \latin{et~al.}(2019)Tian, Zhang, Li, Ying, Li, Zhang, Liu, and Lei]{VI3-2_2019}
Tian,~S.; Zhang,~J.-F.; Li,~C.; Ying,~T.; Li,~S.; Zhang,~X.; Liu,~K.; Lei,~H. Ferromagnetic van der Waals crystal VI3. \emph{Journal of the American Chemical Society} \textbf{2019}, \emph{141}, 5326--5333\relax
\mciteBstWouldAddEndPuncttrue
\mciteSetBstMidEndSepPunct{\mcitedefaultmidpunct}
{\mcitedefaultendpunct}{\mcitedefaultseppunct}\relax
\EndOfBibitem
\bibitem[Hu \latin{et~al.}(2020)Hu, Gong, Zeng, Wang, and Fan]{Fe2Cl3I3_2020}
Hu,~Y.; Gong,~Y.~H.; Zeng,~H.~H.; Wang,~J.~H.; Fan,~X.~L. Two-dimensional stable Fe-based ferromagnetic semiconductors: FeI3 and FeI1.5Cl1.5 monolayers. \emph{Phys. Chem. Chem. Phys.} \textbf{2020}, \emph{22}, 24506--24515\relax
\mciteBstWouldAddEndPuncttrue
\mciteSetBstMidEndSepPunct{\mcitedefaultmidpunct}
{\mcitedefaultendpunct}{\mcitedefaultseppunct}\relax
\EndOfBibitem
\bibitem[Fu \latin{et~al.}(2024)Fu, Li, Bo, Ma, Li, and Pu]{Cr2Cl3I3_2024}
Fu,~L.; Li,~S.; Bo,~X.; Ma,~S.; Li,~F.; Pu,~Y. Two-dimensional Cr2Cl3S3 Janus magnetic semiconductor with large magnetic exchange interaction and high-TC. \emph{Chinese Physics B} \textbf{2024}, \emph{33}, 096301\relax
\mciteBstWouldAddEndPuncttrue
\mciteSetBstMidEndSepPunct{\mcitedefaultmidpunct}
{\mcitedefaultendpunct}{\mcitedefaultseppunct}\relax
\EndOfBibitem
\bibitem[Ren \latin{et~al.}(2020)Ren, Li, Wan, Liu, and Ge]{V2Cl3I3_2020}
Ren,~Y.; Li,~Q.; Wan,~W.; Liu,~Y.; Ge,~Y. High-temperature ferromagnetic semiconductors: Janus monolayer vanadium trihalides. \emph{Physical Review B} \textbf{2020}, \emph{101}, 134421\relax
\mciteBstWouldAddEndPuncttrue
\mciteSetBstMidEndSepPunct{\mcitedefaultmidpunct}
{\mcitedefaultendpunct}{\mcitedefaultseppunct}\relax
\EndOfBibitem
\bibitem[Zhang \latin{et~al.}(2023)Zhang, Liu, Xu, and Gao]{FeClF_2023}
Zhang,~L.; Liu,~Y.; Xu,~Z.; Gao,~G. Electronic phase transition, perpendicular magnetic anisotropy and high Curie temperature in Janus FeClF. \emph{2D Materials} \textbf{2023}, \emph{10}, 045005\relax
\mciteBstWouldAddEndPuncttrue
\mciteSetBstMidEndSepPunct{\mcitedefaultmidpunct}
{\mcitedefaultendpunct}{\mcitedefaultseppunct}\relax
\EndOfBibitem
\bibitem[Li \latin{et~al.}(2017)Li, Wei, Zhao, Huang, and Dai]{Spin_2017}
Li,~F.; Wei,~W.; Zhao,~P.; Huang,~B.; Dai,~Y. Electronic and optical properties of pristine and vertical and lateral heterostructures of Janus MoSSe and WSSe. \emph{The journal of physical chemistry letters} \textbf{2017}, \emph{8}, 5959--5965\relax
\mciteBstWouldAddEndPuncttrue
\mciteSetBstMidEndSepPunct{\mcitedefaultmidpunct}
{\mcitedefaultendpunct}{\mcitedefaultseppunct}\relax
\EndOfBibitem
\bibitem[Jiang and Mi(2023)Jiang, and Mi]{Valley_2023}
Jiang,~J.; Mi,~W. Two-dimensional magnetic Janus monolayers and their van der Waals heterostructures: a review on recent progress. \emph{Mater. Horiz.} \textbf{2023}, \emph{10}, 788--807\relax
\mciteBstWouldAddEndPuncttrue
\mciteSetBstMidEndSepPunct{\mcitedefaultmidpunct}
{\mcitedefaultendpunct}{\mcitedefaultseppunct}\relax
\EndOfBibitem
\bibitem[Tang and Kou(2022)Tang, and Kou]{tang20222d}
Tang,~X.; Kou,~L. 2D Janus transition metal dichalcogenides: Properties and applications. \emph{physica status solidi (b)} \textbf{2022}, \emph{259}, 2100562\relax
\mciteBstWouldAddEndPuncttrue
\mciteSetBstMidEndSepPunct{\mcitedefaultmidpunct}
{\mcitedefaultendpunct}{\mcitedefaultseppunct}\relax
\EndOfBibitem
\bibitem[Liu \latin{et~al.}(2022)Liu, Zheng, Li, Si, Wei, Zhang, and Wang]{liu2022two}
Liu,~P.-F.; Zheng,~F.; Li,~J.; Si,~J.-G.; Wei,~L.; Zhang,~J.; Wang,~B.-T. Two-gap superconductivity in a Janus MoSH monolayer. \emph{Physical Review B} \textbf{2022}, \emph{105}, 245420\relax
\mciteBstWouldAddEndPuncttrue
\mciteSetBstMidEndSepPunct{\mcitedefaultmidpunct}
{\mcitedefaultendpunct}{\mcitedefaultseppunct}\relax
\EndOfBibitem
\bibitem[Ku \latin{et~al.}(2023)Ku, Yan, Si, Zhu, Wang, Wei, Pang, Li, and Zhou]{ku2023ab}
Ku,~R.; Yan,~L.; Si,~J.-G.; Zhu,~S.; Wang,~B.-T.; Wei,~Y.; Pang,~K.; Li,~W.; Zhou,~L. Ab initio investigation of charge density wave and superconductivity in two-dimensional Janus 2 H/1 T-MoSH monolayers. \emph{Physical Review B} \textbf{2023}, \emph{107}, 064508\relax
\mciteBstWouldAddEndPuncttrue
\mciteSetBstMidEndSepPunct{\mcitedefaultmidpunct}
{\mcitedefaultendpunct}{\mcitedefaultseppunct}\relax
\EndOfBibitem
\bibitem[Seeyangnok \latin{et~al.}(2024)Seeyangnok, Ul~Hassan, Pinsook, and Ackland]{seeyangnok2024superconductivity}
Seeyangnok,~J.; Ul~Hassan,~M.~M.; Pinsook,~U.; Ackland,~G. Superconductivity and electron self-energy in tungsten-sulfur-hydride monolayer. \emph{2D Materials} \textbf{2024}, \emph{11}, 025020\relax
\mciteBstWouldAddEndPuncttrue
\mciteSetBstMidEndSepPunct{\mcitedefaultmidpunct}
{\mcitedefaultendpunct}{\mcitedefaultseppunct}\relax
\EndOfBibitem
\bibitem[Li \latin{et~al.}(2024)Li, Wei, Shi, Shi, Si, Liu, and Wang]{li2024machine}
Li,~J.; Wei,~L.; Shi,~X.; Shi,~L.; Si,~J.; Liu,~P.-F.; Wang,~B.-T. Machine learning accelerated discovery of superconducting two-dimensional Janus transition metal sulfhydrates. \emph{Physical Review B} \textbf{2024}, \emph{109}, 174516\relax
\mciteBstWouldAddEndPuncttrue
\mciteSetBstMidEndSepPunct{\mcitedefaultmidpunct}
{\mcitedefaultendpunct}{\mcitedefaultseppunct}\relax
\EndOfBibitem
\bibitem[Ul~Hassan and Pinsook(2024)Ul~Hassan, and Pinsook]{ul2024superconductivity}
Ul~Hassan,~M.~M.; Pinsook,~U. Superconductivity in monolayer Janus Titanium-sulfurhydride (TiSH) at ambient pressure. \emph{Journal of Physics: Condensed Matter} \textbf{2024}, \emph{36}, 325702\relax
\mciteBstWouldAddEndPuncttrue
\mciteSetBstMidEndSepPunct{\mcitedefaultmidpunct}
{\mcitedefaultendpunct}{\mcitedefaultseppunct}\relax
\EndOfBibitem
\bibitem[Seeyangnok \latin{et~al.}(2024)Seeyangnok, Pinsook, and Ackland]{jseeyangPRBWXH2024}
Seeyangnok,~J.; Pinsook,~U.; Ackland,~G.~J. Superconductivity and strain-enhanced phase stability of Janus tungsten chalcogenide hydride monolayers. \emph{Physical Review B} \textbf{2024}, \emph{110}, 195408\relax
\mciteBstWouldAddEndPuncttrue
\mciteSetBstMidEndSepPunct{\mcitedefaultmidpunct}
{\mcitedefaultendpunct}{\mcitedefaultseppunct}\relax
\EndOfBibitem
\bibitem[Seeyangnok \latin{et~al.}(2024)Seeyangnok, Pinsook, and Ackland]{jseeyangnok2024JTMCS}
Seeyangnok,~J.; Pinsook,~U.; Ackland,~G.~J. Superconductivity in Janus IV-B transition metal chalcogenide hydrides. \emph{arXiv preprint arXiv:2410.21769} \textbf{2024}, \relax
\mciteBstWouldAddEndPunctfalse
\mciteSetBstMidEndSepPunct{\mcitedefaultmidpunct}
{}{\mcitedefaultseppunct}\relax
\EndOfBibitem
\bibitem[Qiu \latin{et~al.}(2024)Qiu, Liu, Ge, Cao, Han, and Yang]{CrSH_1T_2024}
Qiu,~X.; Liu,~B.; Ge,~L.; Cao,~L.; Han,~K.; Yang,~H. High Curie temperature ferromagnetic monolayer T-CrSH and valley physics of T-CrSH/WS2 heterostructure. \emph{Phys. Chem. Chem. Phys.} \textbf{2024}, \emph{26}, 5589--5596\relax
\mciteBstWouldAddEndPuncttrue
\mciteSetBstMidEndSepPunct{\mcitedefaultmidpunct}
{\mcitedefaultendpunct}{\mcitedefaultseppunct}\relax
\EndOfBibitem
\bibitem[Jiang \latin{et~al.}(2015)Jiang, Wang, Wang, and Park]{ZA_2015}
Jiang,~J.-W.; Wang,~B.-S.; Wang,~J.-S.; Park,~H.~S. A review on the flexural mode of graphene: lattice dynamics, thermal conduction, thermal expansion, elasticity and nanomechanical resonance. \emph{Journal of Physics: Condensed Matter} \textbf{2015}, \emph{27}, 083001\relax
\mciteBstWouldAddEndPuncttrue
\mciteSetBstMidEndSepPunct{\mcitedefaultmidpunct}
{\mcitedefaultendpunct}{\mcitedefaultseppunct}\relax
\EndOfBibitem
\bibitem[Eriksson \latin{et~al.}(2019)Eriksson, Fransson, and Erhart]{hiphive2019}
Eriksson,~F.; Fransson,~E.; Erhart,~P. The hiphive package for the extraction of high-order force constants by machine learning. \emph{Advanced Theory and Simulations} \textbf{2019}, \emph{2}, 1800184\relax
\mciteBstWouldAddEndPuncttrue
\mciteSetBstMidEndSepPunct{\mcitedefaultmidpunct}
{\mcitedefaultendpunct}{\mcitedefaultseppunct}\relax
\EndOfBibitem
\bibitem[Fransson \latin{et~al.}(2020)Fransson, Eriksson, and Erhart]{hiphive2020}
Fransson,~E.; Eriksson,~F.; Erhart,~P. Efficient construction of linear models in materials modeling and applications to force constant expansions. \emph{npj Computational Materials} \textbf{2020}, \emph{6}, 135\relax
\mciteBstWouldAddEndPuncttrue
\mciteSetBstMidEndSepPunct{\mcitedefaultmidpunct}
{\mcitedefaultendpunct}{\mcitedefaultseppunct}\relax
\EndOfBibitem
\bibitem[Born and Huang(1996)Born, and Huang]{bornhuang1996}
Born,~M.; Huang,~K. \emph{Dynamical theory of crystal lattices}; Oxford university press, 1996\relax
\mciteBstWouldAddEndPuncttrue
\mciteSetBstMidEndSepPunct{\mcitedefaultmidpunct}
{\mcitedefaultendpunct}{\mcitedefaultseppunct}\relax
\EndOfBibitem
\bibitem[{Hohenberg} and {Kohn}(1964){Hohenberg}, and {Kohn}]{DFT1964}
{Hohenberg},~P.; {Kohn},~W. {Inhomogeneous Electron Gas}. \emph{Physical Review} \textbf{1964}, \emph{136}, 864--871\relax
\mciteBstWouldAddEndPuncttrue
\mciteSetBstMidEndSepPunct{\mcitedefaultmidpunct}
{\mcitedefaultendpunct}{\mcitedefaultseppunct}\relax
\EndOfBibitem
\bibitem[Giannozzi \latin{et~al.}(2017)Giannozzi, Andreussi, Brumme, Bunau, Nardelli, Calandra, Car, Cavazzoni, Ceresoli, Cococcioni, Colonna, Carnimeo, Corso, de~Gironcoli, Delugas, Jr, Ferretti, Floris, Fratesi, Fugallo, Gebauer, Gerstmann, Giustino, Gorni, Jia, Kawamura, Ko, Kokalj, Küçükbenli, Lazzeri, Marsili, Marzari, Mauri, Nguyen, Nguyen, de-la Roza, Paulatto, Poncé, Rocca, Sabatini, Santra, Schlipf, Seitsonen, Smogunov, Timrov, Thonhauser, Umari, Vast, Wu, and Baroni]{QE-2017}
Giannozzi,~P. \latin{et~al.}  {Advanced capabilities for materials modelling with QUANTUM ESPRESSO}. \emph{Journal of Physics: Condensed Matter} \textbf{2017}, \emph{29}, 465901\relax
\mciteBstWouldAddEndPuncttrue
\mciteSetBstMidEndSepPunct{\mcitedefaultmidpunct}
{\mcitedefaultendpunct}{\mcitedefaultseppunct}\relax
\EndOfBibitem
\bibitem[Giannozzi \latin{et~al.}(2020)Giannozzi, Baseggio, Bonf{\`a}, Brunato, Car, Carnimeo, Cavazzoni, De~Gironcoli, Delugas, Ferrari~Ruffino, \latin{et~al.} others]{QE-2020}
Giannozzi,~P.; Baseggio,~O.; Bonf{\`a},~P.; Brunato,~D.; Car,~R.; Carnimeo,~I.; Cavazzoni,~C.; De~Gironcoli,~S.; Delugas,~P.; Ferrari~Ruffino,~F.; others {Quantum ESPRESSO toward the exascale}. \emph{The Journal of Chemical Physics} \textbf{2020}, \emph{152}, 154105\relax
\mciteBstWouldAddEndPuncttrue
\mciteSetBstMidEndSepPunct{\mcitedefaultmidpunct}
{\mcitedefaultendpunct}{\mcitedefaultseppunct}\relax
\EndOfBibitem
\bibitem[Cococcioni and De~Gironcoli(2005)Cococcioni, and De~Gironcoli]{DFT_U2005}
Cococcioni,~M.; De~Gironcoli,~S. {Linear response approach to the calculation of the effective interaction parameters in the LDA$+U$ method}. \emph{Physical Review B} \textbf{2005}, \emph{71}, 035105\relax
\mciteBstWouldAddEndPuncttrue
\mciteSetBstMidEndSepPunct{\mcitedefaultmidpunct}
{\mcitedefaultendpunct}{\mcitedefaultseppunct}\relax
\EndOfBibitem
\bibitem[Timrov \latin{et~al.}(2022)Timrov, Marzari, and Cococcioni]{timrov2022hp}
Timrov,~I.; Marzari,~N.; Cococcioni,~M. HP--A code for the calculation of Hubbard parameters using density-functional perturbation theory. \emph{Computer Physics Communications} \textbf{2022}, \emph{279}, 108455\relax
\mciteBstWouldAddEndPuncttrue
\mciteSetBstMidEndSepPunct{\mcitedefaultmidpunct}
{\mcitedefaultendpunct}{\mcitedefaultseppunct}\relax
\EndOfBibitem
\bibitem[Timrov \latin{et~al.}(2018)Timrov, Marzari, and Cococcioni]{timrov2018hubbard}
Timrov,~I.; Marzari,~N.; Cococcioni,~M. Hubbard parameters from density-functional perturbation theory. \emph{Physical Review B} \textbf{2018}, \emph{98}, 085127\relax
\mciteBstWouldAddEndPuncttrue
\mciteSetBstMidEndSepPunct{\mcitedefaultmidpunct}
{\mcitedefaultendpunct}{\mcitedefaultseppunct}\relax
\EndOfBibitem
\bibitem[Timrov \latin{et~al.}(2021)Timrov, Marzari, and Cococcioni]{timrov2021self}
Timrov,~I.; Marzari,~N.; Cococcioni,~M. Self-consistent Hubbard parameters from density-functional perturbation theory in the ultrasoft and projector-augmented wave formulations. \emph{Physical Review B} \textbf{2021}, \emph{103}, 045141\relax
\mciteBstWouldAddEndPuncttrue
\mciteSetBstMidEndSepPunct{\mcitedefaultmidpunct}
{\mcitedefaultendpunct}{\mcitedefaultseppunct}\relax
\EndOfBibitem
\bibitem[Perdew \latin{et~al.}(1996)Perdew, Burke, and Ernzerhof]{perdew1996generalized}
Perdew,~J.~P.; Burke,~K.; Ernzerhof,~M. Generalized gradient approximation made simple. \emph{Physical review letters} \textbf{1996}, \emph{77}, 3865\relax
\mciteBstWouldAddEndPuncttrue
\mciteSetBstMidEndSepPunct{\mcitedefaultmidpunct}
{\mcitedefaultendpunct}{\mcitedefaultseppunct}\relax
\EndOfBibitem
\bibitem[Kresse and Joubert(1999)Kresse, and Joubert]{kresse1999ultrasoft}
Kresse,~G.; Joubert,~D. From ultrasoft pseudopotentials to the projector augmented-wave method. \emph{Physical review b} \textbf{1999}, \emph{59}, 1758\relax
\mciteBstWouldAddEndPuncttrue
\mciteSetBstMidEndSepPunct{\mcitedefaultmidpunct}
{\mcitedefaultendpunct}{\mcitedefaultseppunct}\relax
\EndOfBibitem
\bibitem[Hobbs \latin{et~al.}(2000)Hobbs, Kresse, and Hafner]{hobbs2000fully}
Hobbs,~D.; Kresse,~G.; Hafner,~J. {Fully unconstrained noncollinear magnetism within the projector augmented-wave method}. \emph{Physical Review B} \textbf{2000}, \emph{62}, 11556\relax
\mciteBstWouldAddEndPuncttrue
\mciteSetBstMidEndSepPunct{\mcitedefaultmidpunct}
{\mcitedefaultendpunct}{\mcitedefaultseppunct}\relax
\EndOfBibitem
\bibitem[Monkhorst and Pack(1976)Monkhorst, and Pack]{monkhorst1976special}
Monkhorst,~H.~J.; Pack,~J.~D. Special points for Brillouin-zone integrations. \emph{Physical review B} \textbf{1976}, \emph{13}, 5188\relax
\mciteBstWouldAddEndPuncttrue
\mciteSetBstMidEndSepPunct{\mcitedefaultmidpunct}
{\mcitedefaultendpunct}{\mcitedefaultseppunct}\relax
\EndOfBibitem
\bibitem[Togo \latin{et~al.}(2023)Togo, Chaput, Tadano, and Tanaka]{phonopy-phono3py-JPCM}
Togo,~A.; Chaput,~L.; Tadano,~T.; Tanaka,~I. Implementation strategies in phonopy and phono3py. \emph{J. Phys. Condens. Matter} \textbf{2023}, \emph{35}, 353001\relax
\mciteBstWouldAddEndPuncttrue
\mciteSetBstMidEndSepPunct{\mcitedefaultmidpunct}
{\mcitedefaultendpunct}{\mcitedefaultseppunct}\relax
\EndOfBibitem
\end{mcitethebibliography}

\end{document}